\documentclass[12pt]{article}

\usepackage{epsf,epsfig,epstopdf}
\usepackage{graphics}
\usepackage{cite}
\usepackage{amsmath,amssymb}
\usepackage{url}
\usepackage{a4wide}

\newcommand{\dirac}[1]{\displaystyle{\not} #1}

\newcommand{\Eqn}[1]{Eq.~(\ref{#1})}
\newcommand{\Eqns}[2]{Eqs.~(\ref{#1}) and (\ref{#2})}

\numberwithin{equation}{section}
  \newcommand{\ccaption}[2]{
    \begin{center}
    \parbox{0.95\textwidth}{
      \caption[#1]{\small{#2}}
      }
    \end{center}
    }

\begin{document}
\thispagestyle{empty}
\allowdisplaybreaks

\begin{flushright}
{\small
IPPP/10/52}\\
July 2010
\end{flushright}

\vspace{\baselineskip}

\begin{center}

\vspace{0.5\baselineskip} \textbf{\Large\boldmath
Production-decay interferences at NLO in QCD for \\[8pt]
$t$-channel single-top production
}
\\
\vspace{3\baselineskip}
{\sc P.~Falgari, P.~Mellor, A.~Signer}\\
\vspace{0.7cm}
{\sl IPPP, Department of Physics, University of Durham, \\
Durham DH1 3LE, England}

\vspace*{1.2cm}

\textbf{Abstract}\\

\vspace{1\baselineskip}

\parbox{0.9\textwidth}{We present a calculation of ${\cal O}(\alpha_s)$
  contributions to the process of $t$-channel single-top production
  and decay, which include virtual and real corrections arising from
  interference of the production and decay subprocesses. The
  calculation is organized as a simultaneous expansion of the matrix
  elements in the couplings $\alpha_{ew},\alpha_s$ and the virtuality
  of the intermediate top quark, $(p_t^2-m_t^2)/m_t^2 \sim
  \Gamma_t/m_t$, and extends earlier results beyond the narrow-width
  approximation.}

\end{center}

\newpage
\setcounter{page}{1}


\section{Introduction}

Both the D0 and CDF collaborations have recently announced the
observation of single-top production at the Fermilab Tevatron at a
significance of $5$ standard deviations \cite{Aaltonen:2009jj,
  Abazov:2009ii}. This process represents a promising channel for the
study of the charged-current weak interactions of the top quark, and
will play a prominent role in the physics program at the LHC, where
top quarks will be produced singly at large rates. Measurements of the
single-top production cross section can be used to directly determine
$V_{t b}$ and to test the unitarity of the CKM matrix
\cite{Alwall:2006bx}. Furthermore, angular correlations of the
products of the top-quark decay encode information on the spin
structure of the $W t b$ vertex and on the production dynamics of the
top quark \cite{Mahlon:1999gz, Motylinski:2009kt}. The single-top
production cross section also probes the bottom-quark distribution
inside the proton, which at the moment is computed from light-parton
densities rather than extracted from data. Therefore, this reaction
represents a means of directly constraining the heavy-quark content of
the proton. Finally, single-top production constitutes a background to
a number of possible new-physics processes, most notably some channels
important for Higgs boson searches. In view of all these
considerations, a precise theoretical description of single-top
production in hadronic collisions is highly desirable.

In the Standard Model (SM) single-top production proceeds via three
main hadronic channels, namely, $t$-channel production, $q b \rightarrow
q' t$ or $\bar{q} b \rightarrow \bar{q}' t$; $s$-channel production, $q
\bar{q}' \rightarrow t \bar{b}$; and associated $t W$ production, $b g
\rightarrow W^- t$.  At the Tevatron and at the LHC the $t$-channel
production process has the largest cross section. In particular, at
the LHC the $s$-channel production cross section and the $t W$
production cross section are expected to be respectively 20 times and
3 times smaller than the $t$-channel process (see
e.g. Ref.~\cite{Bernreuther:2008ju}).  Thus, in this paper we
concentrate on the $t$-channel production mechanism. However, the
distinction between $t$-channel and $s$-channel production is
problematic beyond leading order and we will have to be more precise
(in Section~\ref{sec:method}) in describing how exactly we construct
our observables and what we include in our calculation.

Top-quark production, or the production of any unstable heavy
particle, can be treated in several ways. The most straightforward way
is to treat the top as a stable particle and ignore its decay. In this
context, the cross section for the $t$-channel single-top production to
next-to-leading order (NLO) in QCD was computed in
Refs.~\cite{Bordes:1994ki, Stelzer:1997ns}. Later, a fully differential
calculation was presented in Refs.~\cite{Harris:2002md,
  Sullivan:2004ie}. NLO QCD corrections have also been computed for
$s$-channel and associated $t W$ production~\cite{Smith:1996ij,
  Harris:2002md, Giele:1995kr}.  The full electroweak corrections in
the SM and MSSM were computed in Ref.~\cite{Beccaria:2008av} for
stable $t$-channel single-top production, and very recently for both
$t$- and $s$-channel processes \cite{Macorini:2010bp}.  Finally,
effects of soft-gluon corrections beyond NLO have also been
studied~\cite{Kidonakis:2007ej}.

Beyond the stable-top approximation, the one-loop corrections split in
a gauge-invariant way into so-called \emph{factorizable} and
\emph{non-factorizable} corrections~\cite{Denner:1997ia,
  Beenakker:1997ir}. Factorizable corrections correspond to (on-shell)
corrections to either the production or the decay part of the
process. Thus, a way to simplify the calculation is to separately
compute the corrections to the production and decay of an on-shell
top. This approximation (sometimes referred to as the improved
narrow-width approximation) allows the inclusion of realistic cuts on
the decay products of the top. NLO QCD analyses in this framework for
the semileptonic top decay were published in
Refs.~\cite{Campbell:2004ch, Cao:2004ky,Cao:2005pq, Campbell:2005bb}.

To our knowledge, none of the presently published works on $t$-channel
single-top production include the effects of interference between real
radiation in production and decay or virtual corrections connecting
the two subprocesses. A study of these non-factorizable
contributions has been presented for $s$-channel single-top
production~\cite{Pittau:1996rp}, and for $t \bar{t}$
production~\cite{Beenakker:1999ya}. These corrections are known to be
very small, for observables which are inclusive enough in the invariant
mass of the top quark \cite{Fadin:1993dz, Melnikov:1993np}, due to
large cancellations between virtual and real contributions. However,
there is, a priori, no reason why this should hold true for arbitrary
observables, especially if they involve kinematical cuts that could, in
principle, spoil the delicate cancellation of real and virtual
contributions. 

In this paper we want to assess the effect of these production-decay
interferences in $t$-channel single-top production at NLO in the QCD
coupling constant, $\alpha_s$. Hence, we are interested in resonant
top-quark production with $p_t^2 \neq m_t^2$, but $p_t^2 \simeq
m_t^2$ or, more precisely, $p_t^2-m_t^2 \sim m_t \Gamma_t$, where
$\Gamma_t \simeq 1.4$~GeV is the width of the top. While the effect of
these ``off-shell'' corrections is expected to be very modest for the
total cross section, we are particularly interested in distributions
that are related to the measurement of the top-quark mass, $m_t$. For
a reliable mass determination with an error $\delta m_t < \Gamma_t$,
the non-factorizable corrections have to be under control. In this
respect, we also view the current calculation as a preparation to
apply our method to top-quark pair production.

We will neglect quark-mixing effects and treat the bottom quark as
massless throughout, using the 5-flavour scheme. The importance of
bottom-quark mass corrections and the relation between the 4-flavour
and 5-flavour scheme has been studied in~Refs.~\cite{Campbell:2009ss,
  Campbell:2009gj}. Furthermore, our calculation does not include any
effects due to parton showers. The matching of the NLO QCD result with
parton shower Monte Carlo was implemented in
MC@NLO~\cite{Frixione:2005vw} and in POWHEG~\cite{Alioli:2009je}.

The calculation is organized as an expansion in the virtuality of the
top quark, $p_t^2-m_t^2$, in a way similar to the pole approximation
\cite{Stuart:1991xk, Aeppli:1993rs}, and employs effective-theory (ET)
inspired techniques analogous to the ones used in
Refs.~\cite{Beneke:2003xh, Beneke:2004km, Beneke:2007zg}. These are
based on splitting contributions to the amplitude into so-called hard
and soft parts using the method of regions~\cite{Beneke:1997zp,
  Smirnov:2002pj}, thereby extending the separation between
factorizable and non-factorizable corrections beyond NLO. The hard
part can be identified with the factorizable corrections, whereas the
soft part corresponds to the non-factorizable
contributions~\cite{Chapovsky:2001zt}. This approach has the advantage
of providing a gauge-invariant resummation of top-quark finite-width
effects.  Furthermore, the expansion in $p_t^2-m_t^2$ allows for an
identification of the terms relevant to the achievement of a given
target accuracy before the actual computation, leading to a
significant simplification of the calculation. The method has been
discussed in detail for a toy model~\cite{Beneke:2004km} and can
easily be adapted to our case for the tree-level and virtual
contributions. For the real corrections this is more problematic and
we will not be able to follow a strict ET approach in this case, but
will have to combine ideas from the effective theory with a standard
fixed-order approach.  As will be shown in this paper, this results in
straightforward calculations of the contributions that are expected to
be relevant for phenomenological applications at hadron colliders.

The outline of the paper is as follows: we start with a general
description of our method to deal with resonant particles at hadron
colliders in Section~\ref{sec:method}. While we will concentrate on
the process at hand, the discussion is meant to be general enough to
be easily adapted to other processes. We will also be more precise in
describing the observables we are interested in and the accuracy we
are aiming for. In Section~\ref{sec:amplitudes} we first give explicit
results for the amplitudes needed and details of the computation. We
then discuss a series of successive approximations to the exact 
cross section, which relate to previous results in the literature, 
and illustrate the
cancellation of infrared singularities in the various
cases. Numerical results for the cross sections and
distributions 
will be presented in Section~\ref{sec:results}. Finally,
in Section~\ref{sec:conclusion} we summarize and give an outlook on
further possible applications of our method.

\section{Method}
\label{sec:method}

\subsection{Setup of the calculation}
\label{sec:setup}

In order to include production-decay interference effects, the
narrow-width approximation has to be relaxed. In particular, the top
quark cannot be treated as a stable particle. Physical
observables must be computed for final states containing only
long-enough lived particles, which in the case of interest in this
paper are represented by the products of the top-quark
decay. Considering the LHC, we are interested in the process
\begin{equation} \label{eq:lhc_process}
p(P_1)\, p(P_2) \to J_b(p_b)\, W^+(p_W)\, X \, ,
\end{equation}
where $J_b$ is a $b$-quark jet and $X$ stands for an arbitrary number
of further jets, as long as they do not originate from a $b$ or $\bar{b}$
quark. Rejecting $\bar{b}$-quark jets (na\"{\i}vely) excludes contributions
from the $s$-channel process. Furthermore, we do not allow a second
$W$ in the final state in order to exclude associated production, and we
insist on a positively charged lepton (from the $W$ decay) to exclude
single-$\bar{t}$ production. Some of these constraints are, of course,
questionable from an experimental point of view and most could easily
be avoided. But this is a minimal, more or less realistic setup that
allows us to discuss the inclusion of non-factorizable corrections.

The most important constraint we make on the final state is that the
invariant mass of the $W^+$$J_b$ pair is close to the top-quark mass,
i.e. that the top is resonant. More precisely, we require\footnote{If
  taken at face value this would correspond to an invariant-mass
  window $\Delta p_t \sim \Gamma_t$ around the top-quark mass. In
  fact, a numerical study reveals that the suppression of non-resonant
  configurations is already effective for much looser invariant-mass
  cuts.}  $(p_b+p_W)^2 - m_t^2 \sim m_t \Gamma_t \sim \alpha_{ew}
m_t^2 \ll m_t^2$ . As is well known, in this case strict fixed-order
perturbation theory breaks down due to the kinematic enhancement of
formally higher-order corrections. These corrections have to be
resummed in a consistent way.

To illustrate this, consider the process (\ref{eq:lhc_process}) at partonic 
tree level, where we have to compute
\begin{equation} \label{eq:qWb}
q(p_1) \, b(p_2) \rightarrow
q'(p_3) \, b(p_4) \, W^+(p_W) \rightarrow
q'(p_3) \, b(p_4) \,l^+(p_5) \, \nu_l(p_6) \, ,
\end{equation}
with the initial parton, $q$, being a light quark ($u, \, c$) or
antiquark ($\bar{d}, \, \bar{s}$). Accordingly, $q'$ is either a quark
($d, \, s$) or antiquark ($\bar{u}, \, \bar{c}$) respectively. For the
purpose of the current discussion the $b$ quark can be identified with
a $b$ jet, $J_b$, with momentum $p_{b}=p_4$. The decay of the $W^+ \to
l^+ \nu_l$ is described in the improved narrow-width approximation. 
This considerably simplifies the calculation,
and yet allows for a non-inclusive treatment of the leptons coming
from the $W$-boson decay.

\begin{figure}[t!]
\begin{center}
\includegraphics[width=0.9 \linewidth]{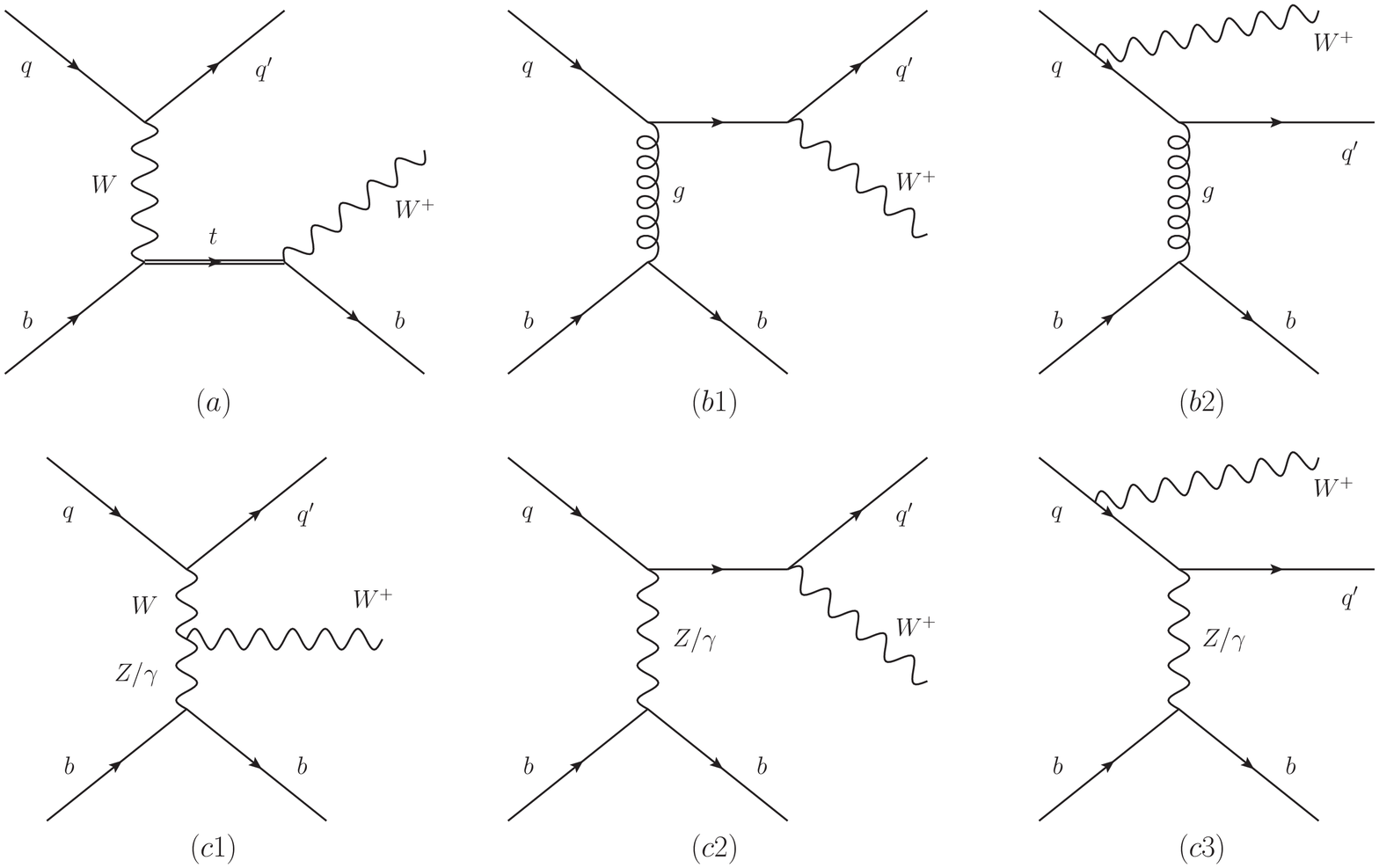}
\end{center}
\ccaption{}{Tree-level Feynman diagrams for the process $q b \rightarrow
  q' b W$. The figure shows both purely EW contributions, diagrams (a) and (c),
  and mixed QCD-EW contributions, diagrams (b). The semileptonic
  (on-shell) decay of the $W$ is understood.\label{fig:qWbtree}}
\end{figure}

The matrix element for the partonic process (\ref{eq:qWb}) can be
computed from the Feynman diagrams shown in
Figure~\ref{fig:qWbtree}. These can be divided into two classes:
\emph{resonant diagrams} (diagram (a)), in which the $W b$ pair
originates from the decay of an internal top quark; and
\emph{non-resonant} or \emph{background diagrams} (diagrams (b) and
(c)), that do not contain intermediate top-quark lines. The latter are
subdivided into electroweak-mediated (diagrams (c)) and mixed QCD-EW
diagrams (diagrams (b)). It is important to note that only the sum
of all electroweak diagrams (diagrams (a) and (c)) is gauge invariant,
though, strictly speaking, only diagram (a) describes the production
of a single top quark. Obviously, the sum of all QCD-EW diagrams is
separately gauge independent.

The fixed-order tree-level amplitude, ${\cal A}^{\text{tree}}$,
can be written as
\begin{equation} \label{eq:AmpFO}
{\cal A}^{\text{tree}} = 
\frac{{\cal K}(p_i)}{p_t^2-m_t^2}
+ {\cal J}(p_i) \, ,
\end{equation}
where ${\cal K}(p_i)$ and ${\cal J}(p_i)$ are functions of the external
momenta, $p_i, \, i=1,...,6$. The first term in \Eqn{eq:AmpFO}
describes resonant contributions, whereas ${\cal J}$ accounts for
non-resonant diagrams. \Eqn{eq:AmpFO} has a pole at $p_t^2=m_t^2$
that is related to the breakdown of fixed-order perturbation theory 
mentioned above.  When an intermediate virtual top quark approaches
the mass shell, a subset of formally subleading corrections to the
top-quark propagator are enhanced, and must be resummed to all orders:
\begin{equation}
\frac{i (\dirac{p_t}+m_t)}{p_t^2-m_t^2} \rightarrow 
\frac{i (\dirac{p_t}+m_t)}{p_t^2-m_t^2}
\sum_{n=0}^{\infty}
\left[-i \bar{\Sigma}_t(\dirac{p_t}) 
\frac{i (\dirac{p_t}+m_t)}{p_t^2-m_t^2} \right]^n \, .
\label{prop_resum}
\end{equation}
$\bar{\Sigma}_t(\dirac{p_t})$ denotes the sum of (renormalized)
one-particle irreducible corrections to the top-quark two-point
function, and contains an imaginary part of order $\alpha_{ew} m_t$
that, upon resummation, regularizes the propagator. However, mixing
different orders in perturbation theory leads, in general, to
violation of gauge invariance and unitarity, which are guaranteed only
for strictly fixed-order calculations and for the full
amplitude. Therefore, a meaningful gauge-invariant expansion of the matrix
element in $p_t^2-m_t^2$ requires addressing the issue of resummation
of finite-width effects.

In recent years, several frameworks for the gauge-invariant
calculation of production cross sections for unstable particles have
been proposed, from ad-hoc procedures to treat weak gauge bosons
\cite{Argyres:1995ym}, to more general methods based on complex
renormalization \cite{Denner:1999gp,Denner:2006ic} or modified
perturbative expansion in terms of distributions
\cite{Tkachov:1998uy,Nekrasov:1999cj,Nekrasov:2002mw}.  Here we adopt
an effective-theory inspired approach~\cite{Beneke:2004km} which, in
this case, can be considered as an extension of the pole-approximation
scheme~\cite{Stuart:1991xk, Aeppli:1993rs}. As we will see, this
provides a consistent framework for a systematic gauge-invariant
expansion of the amplitude around the complex pole  
of the full top propagator at $\mu_t^2 \equiv
m_t^2-i m_t \Gamma_t$.  Note that we have
used the same notation, $m_t$, for the mass parameter in
\Eqn{prop_resum} and for the real part of the complex pole.  Strictly
speaking, the first one is the scheme-dependent renormalized mass,
$m_{t,r}$, and the latter the physical pole mass\footnote{It is well
  known that for a mass determination with an error $\delta m_t
  \lesssim \Lambda_{\text{QCD}}$, a mass definition other than the
  pole mass has to be used~\cite{Smith:1996xz,
    Fleming:2007qr}. However, in this paper we ignore effects
  of the order $\Lambda_{\text{QCD}}\ll m_t \delta$.}, and in a
generic renormalization scheme they differ by an amount proportional
to $\alpha_s$, as shown in Section~\ref{sec:ren}.  However, in the
on-shell scheme adopted in this paper, the two masses coincide up to
corrections which are beyond our target accuracy. Therefore, in the
following we refrain from introducing more than one mass parameter,
except for Section~\ref{sec:ren}, where it will be necessary to
explicitly distinguish the renormalized mass, $m_{t,r}$, from the pole
mass, $m_t$.

For the tree-level amplitude, \Eqn{eq:AmpFO}, the pole expansion reads
\begin{equation} \label{eq:DPA}
{\cal A}^{\text{tree}} = \frac{{\cal K}(p_i;p_t^2=\mu_t^2)}{\Delta_t}
\left(1+\delta R_t\right)
+\frac{\partial {\cal K}}{\partial p_t^2}
(p_i,p_t^2=\mu_t^2)+
{\cal J}(p_i;p_t^2=\mu_t^2)+... \, ,
\end{equation}
where we have introduced $\Delta_t \equiv p_t^2-\mu_t^2$, 
and $1+\delta R_t$ denotes the residue of the full propagator
at $p_t^2=\mu_t^2$. The leading resonant contribution is encoded
in ${\cal K}(p_i;p_t^2=\mu_t^2)$. In $\partial {\cal K}/\partial
p_t^2$ the resonant propagator has been cancelled by a $\Delta_t$
arising from the expansion of ${\cal K}(p_i)$ around $p_t^2=\mu_t^2$,
while ${\cal J}(p_i;p_t^2=\mu_t^2)$ represents the leading
contribution of the truly non-resonant diagrams. The ellipsis denotes
higher-order contributions suppressed by additional powers of
$\Delta_t$.  These can, in principle, be computed to any order, but in
the following discussion only the terms shown in \Eqn{eq:DPA} will be
relevant. 

Beyond the leading approximation, ${\cal A}^{\text{tree}} \sim {\cal
  K}(p_i;p_t^2=\mu_t^2)/\Delta_t$, the consistency of the expansion
(\ref{eq:DPA}) requires the inclusion of loop-corrections.  The
kinematic expansion in the parameter, $\Delta_t$, must be combined with
a standard expansion in the coupling constants, $\alpha_s =
g_s^2/(4\pi)$ and $\alpha_{ew} = g_{ew}^2/(4\pi)$. For counting
purposes, in the following we refer to the three expansions
parameters collectively as $\delta$ and assume the relative scaling
\begin{equation}\label{eq:scaling}
\delta \sim \alpha_s^2 \sim \alpha_{ew} \sim \frac{\Delta_t}{m^2_t} \, .
\end{equation}
We are thus led to write the tree-level amplitude for the process
(\ref{eq:qWb}) as
\begin{equation}
{\cal A}^{{\rm tree}} = \delta_{31} \delta_{42} \Big(  
g_{ew}^3\, A^{(3,0)}_{(-1)} + g_{ew}^3\, A^{(3,0)}_{(0)} +
\ldots \Big) +
T^a_{31} T^a_{42}\,g_{ew} g_s^2\, A^{(1,2)} \, .
\label{treeA}
\end{equation}
The superscripts denote the order of the couplings which multiply the
amplitude, while the subscripts denote the order to which the
propagator, $\Delta_t$, appears within the amplitude,
i.e. $A^{(m,n)}_{(l)}$ has a prefactor $g_{ew}^m\, g_s^n$ with
$A^{(m,n)}_{(l)} \sim \Delta_t^{l}$.  The coupling from the decay of
the $W$ is not counted. A missing subscript indicates that the
amplitude does not contain a $\Delta_t$ propagator, i.e. $A^{(m,n)}
\equiv A^{(m,n)}_{(0)}$. Thus, $g_{ew}^3\, A^{(3,0)}_{(-1)} \sim
\delta^{1/2}$, $g_{ew} g_s^2\, A^{(1,2)} \sim \delta$ and $g_{ew}^3\,
A^{(3,0)}_{(0)} \sim \delta^{3/2}$. Terms suppressed beyond
$\delta^{3/2}$ are indicated by the ellipsis. In the notation of
\Eqn{eq:DPA}, $g_{ew}^3 A^{(3,0)}_{(-1)}$ corresponds to ${\cal
  K}(p_i;p_t^2=\mu_t^2)/\Delta_t$, while $g_{ew}^3 A^{(3,0)}_{(0)}$
includes both the $\partial {\cal K}/\partial p_t^2$ and ${\cal J}$
terms. Clearly, $g_{ew} g_s^2\, A^{(1,2)}$ receives contributions from
non-resonant diagrams only.

Note that the QCD-EW contribution, $A^{(1,2)}$, is
usually considered to be a background to single-top
production. However, the final state is identical and, therefore,
$A^{(1,2)}$ has to be included in ${\cal A}^{{\rm tree}}$. In
principle, there could be interferences between $A^{(3,0)}_{(-1)}$
and $A^{(1,2)}$, however, due to colour they vanish at tree level.
For the colour-averaged squared amplitude, $M^{{\rm tree}} =
\frac{1}{N_c^2} \sum_{c}|{\cal A}^{{\rm tree}}|^2$, we obtain
\begin{equation}
M^{{\rm tree}} = 
g_{ew}^6\,
  \left|A^{(3,0)}_{(-1)}\right|^2
+ g_{ew}^6\,   
2\, {\rm Re}\left(A^{(3,0)}_{(-1)}\,[A^{(3,0)}_{(0)}]^*\right)
+ g_{ew}^2 g_s^4\, \frac{C_F}{2 N_c} 
\left|A^{(1,2)}\right|^2 + \ldots
\label{treeM}
\end{equation}
where, as usual, $N_c=3$ and $C_F = 4/3$. The (first) leading term of
\Eqn{treeM} scales as $\delta$, whereas the other two terms scale
as $\delta^2$ and represent a correction of order $\delta \sim 1 \%$ to the
leading contribution. All other terms are further suppressed.

Our aim is to compute all contributions to the cross section up to
${\cal O}(\delta^{3/2})$.  According to our counting,
\Eqn{eq:scaling}, this requires, beside the leading tree-level
amplitude, the calculation of $A^{(3,2)}_{(-1)}$, the ${\cal
  O}(\alpha_s)$ one-loop corrections to the leading resonant
contribution. For the squared amplitude this leads to a contribution
\begin{equation}
M^{{\rm NLO}} = g_{ew}^6\,  g_s^2\, 
2\, {\rm Re} \left(A^{(3,2)}_{(-1)}\,[A^{(3,0)}_{(-1)}]^*\right)\, ,
\label{nloM}
\end{equation}  
which we will refer to as NLO. Along with the virtual corrections we
also have to include real corrections, $q\,b\to q'\,b\,W^+\,g$. There
are also gluon-initiated processes at NLO, namely $g b\to
q'\,b\,W^+\,\bar{q}$ and $q\,g\to q'\,b\,W^+\,\bar{b}$. The former is
fully included in our calculation, whereas the latter deserves special
discussion as it mixes $t$-channel and $s$-channel single-top
production. Writing the amplitude for $q(p_1)\,g(p_2)\to
q'(p_3)\,b(p_4)\,W^+(p_W)\, \bar{b}(p_7)$ as
\begin{equation}
{\cal A}^{{\rm tree}}_{qg} = 
g_s\, g_{ew}^3\, \Big(
T^{a_2}_{47} \delta_{31}\,A^{47}_{qg} +
T^{a_2}_{31} \delta_{47}\,A^{31}_{qg} \Big) \, ,
\label{treeAqg}
\end{equation}
we first note that upon squaring ${\cal A}^{{\rm tree}}_{qg} $ there are no
interference terms, due to colour. Conventionally, the term $\sim
|A^{47}_{qg}|^2$ is included in $t$-channel single-top production,
whereas the term $\sim |A^{31}_{qg}|^2$ is considered to be a higher-order
correction to $s$-channel single-top production. We will follow this
convention but stress that a fully satisfactory solution requires the
simultaneous inclusion of both processes, which we reserve for
future work.

In summary, we need to compute the tree-level amplitudes for
\begin{eqnarray}
&u\,b \to d\, b\, W^+ \phantom{\, g} 
    &\qquad  \bar{d}\,b \to \bar{u}\, b\, W^+
\label{process:ub}
\\
& u\,b \to d\, b\, W^+\, g 
    &\qquad  \bar{d}\,b \to \bar{u}\, b\, W^+\, g
\label{process:ubg}
\\
& u\,g \to d\, b\, W^+\, \bar{b}
    &\qquad  \bar{d}\,g \to \bar{u}\, b\, W^+\, \bar{b}
\label{process:ugb}
\\
& g\,b \to d\, b\, W^+\, \bar{u} & 
\label{process:gbu}
\end{eqnarray}
and the QCD one-loop corrections to the leading (in $\Delta_t/m_t$)
part of process (\ref{process:ub}). Of course, we will also have to
include the processes with $\{u,d\} \to \{c,s\}$. Note that the
various processes containing a gluon are related by crossing.  We
stress once more that the semileptonic decay, $W^+\to \ell^+ \nu_\ell$,
is taken into account in the improved narrow-width approximation.

\subsection{Loop corrections in the effective-theory approach}
\label{sec:et}

Detailed discussions and applications of effective-theory methods to
the calculation of processes involving unstable particles can be found
in the literature~\cite{Chapovsky:2001zt, Beneke:2003xh,
  Beneke:2004km, Falgari:2009zz}. Here we will restrict ourselves to a
discussion of the points directly relevant to our calculation.

The main idea of the ET approach is to systematically exploit the
hierarchy of scales, $(p_t^2-m_t^2) \sim m_t \Gamma_t \ll m^2_t$, by
integrating out degrees of freedom with virtuality $\sim m_t^2$. In
doing this, the Lagrangian of the underlying theory is rewritten as a
series of gauge-invariant operators multiplied by matching
coefficients, which are determined such that the ET reproduces the
results of the underlying theory up to a certain approximation. The
matching coefficients are guaranteed to be gauge independent and they
contain the information on the degrees of freedom that have been
integrated out.

Once the hard part is integrated out, we are left with degrees of
freedom with virtuality much smaller than $m_t^2$. In general, we have
to take into account several different such degrees of freedom, but in
the case at hand it is sufficient (to the accuracy we are aiming at)
to consider only soft modes. 

Within this picture, the leading contribution to the
process~(\ref{eq:qWb}) is to be viewed as the production of an
on-shell top, the propagation of a soft top, and the subsequent decay
of an on-shell top. This is to say that the matching of the full
theory onto the ET is done on-shell. It is important to note that
on-shell in this context means $p_t^2 = \mu_t^2$, i.e. on the complex
pole of the two-point function. Thus, the matching coefficients are
generally complex. The propagator of the soft top is derived from the
bilinear operator in the effective theory and includes the resummation
of the self-energy insertions.

There are several sources of higher-order corrections. First, there
are ${\cal O}(\alpha_s)$ and ${\cal O}(\alpha_{ew})$ corrections to
the matching coefficients. These correspond to loop corrections to the
production and decay of a stable top and have been taken into account
in previous calculations. Furthermore, there are subleading
corrections to the bilinear operator, i.e. corrections to the
propagator. There are also subleading operators in the ET that do not
refer to a top quark at all. These operators reproduce the effect of
background diagrams, as well as subleading effects of resonant
diagrams, and correspond to the terms ${\cal J}$ and $\partial {\cal
  K}/\partial p_t^2$ in \Eqn{eq:DPA} respectively.  Finally, there are
the contributions from loop diagrams in the effective theory,
i.e. loop diagrams with the still dynamical soft degrees of
freedom. These contributions correspond to the non-factorizable
corrections and link the production and decay parts of the process.

From a technical point of view, the calculation of the aforementioned 
corrections is achieved by
using the method of regions~\cite{Beneke:1997zp, Smirnov:2002pj} and
computing the hard and soft parts of loop integrals. The hard part is defined by
expanding the integrand under the assumption that the loop momentum,
$q^\mu$, scales as $q\sim m_t$. In what follows we will always suppress
powers of $m_t$ in the scaling relations and write the scaling of the
hard modes as $q^\mu \sim 1$. 
The soft part is obtained by 
expanding the integrand of a loop integral assuming that the loop 
momentum scales as $q \sim \delta$.

A strict application of the effective theory would require the
introduction of collinear fields for the massless external fermions
and a heavy-quark field for the top quark, as well as soft and
collinear gauge-boson fields. However, we refrain from doing this
because for the real corrections we will have to deviate from a strict
ET approach anyway. We will, however, make extensive use of the
gauge-invariant separation of the one-loop contribution into hard and
soft parts, and the associated counting rules to obtain a result that
is gauge invariant and reproduces all terms to the desired accuracy
with a minimal amount of computation.

\begin{figure}[t!]
\begin{center}
\includegraphics[width=0.9 \linewidth]{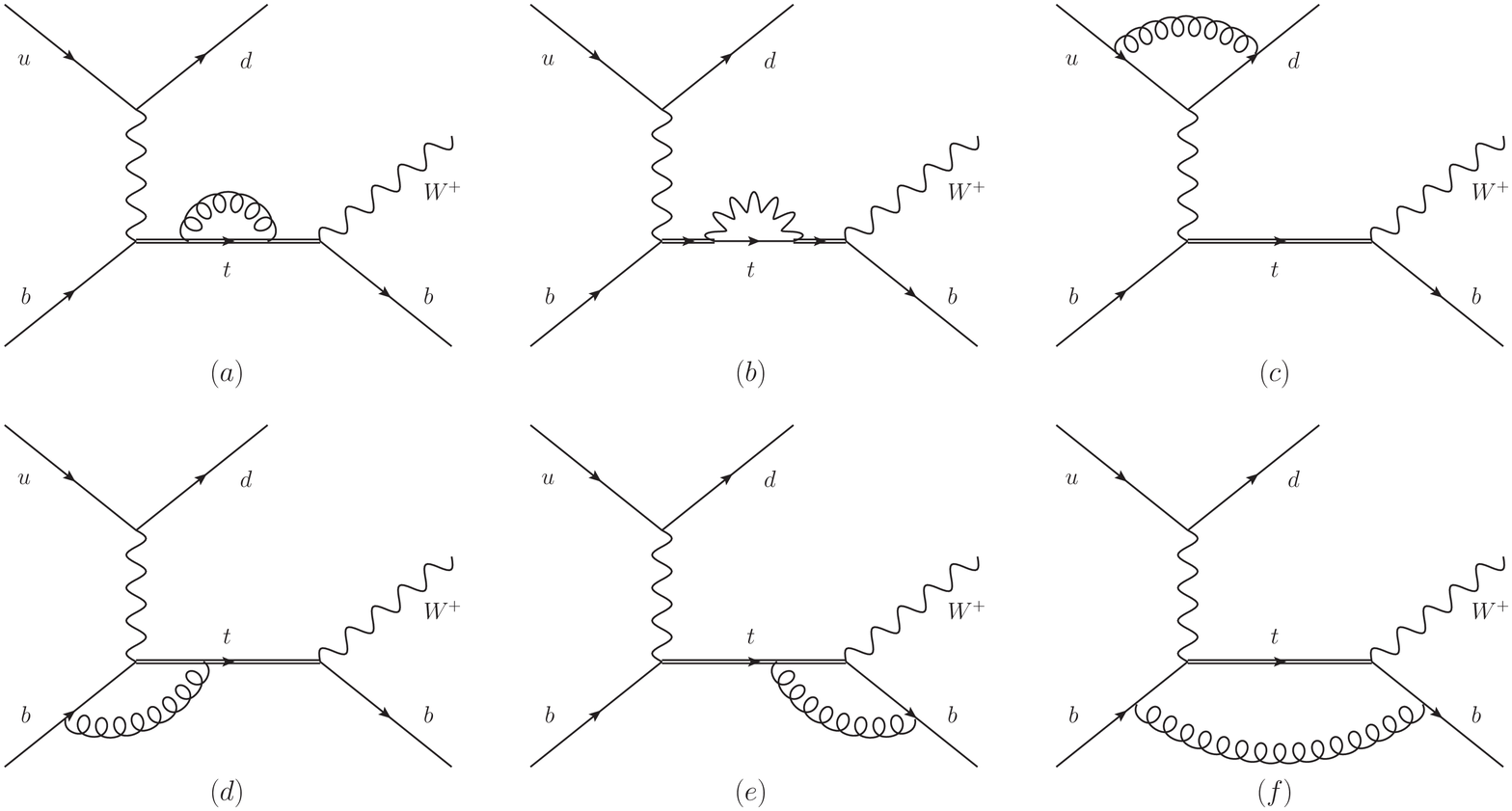}
\end{center}
\ccaption{}{Virtual QCD corrections to t-channel single-top production
  at leading order in $\Delta_t/m_t$.\label{fig:1loop}}
\end{figure}

The one-loop diagrams necessary to achieving ${\cal O} (\delta^{3/2})$
accuracy are shown in Figure~\ref{fig:1loop}. Apart from the top
self-energy insertion (diagram~(\ref{fig:1loop}b)), at the accuracy we are
interested in we have to deal with QCD one-loop diagrams only. Due to
colour, the one-loop diagrams containing gluon exchange between the upper
and lower quark lines result in a vanishing contribution to the
amplitude squared and are therefore not shown in the figure.  A
detailed discussion of the computation of the various diagrams will be
given in Section~\ref{sec:virtual}. Here we briefly mention which
parts of the diagrams in Figure~\ref{fig:1loop} enter our calculation.

As explained in detail in Section~\ref{sec:virtual}, taking the
leading hard and soft parts of the QCD self-energy insertion
(diagram~(2a)) we obtain the scalings $g_{ew}^3\alpha_s\,
\delta^{-2}\sim 1$ and $g_{ew}^3\alpha_s\, \delta^{-1}\sim \delta$
respectively.  For the soft part this corresponds precisely to a
correction we are interested in. The hard part, however, seems to be
superleading, i.e. the one-loop correction is enhanced compared to the
tree-level amplitude (which scales as $g_{ew}^3/\delta \sim
\delta^{1/2}$).  However, as we will see below, in the on-shell scheme
used in this paper this superleading contribution is cancelled
precisely by the counterterm. In a generic renormalization scheme this
is not the case, and the leading hard part of the self-energy has to
be resummed and enters the definition of the complex pole, $\mu_t^2$,
as explained in detail in Section~\ref{sec:ren}. The same applies to
the two-loop QCD self-energy. An insertion of the leading hard part
results in a diagram scaling as $g_{ew}^3\alpha^2_s\, \delta^{-2}\sim
\delta^{1/2}$ and, in general, has to be resummed.  The subleading hard
terms are suppressed by at least one factor $p_t^2-m_t^2\sim \delta$
compared to the leading hard contribution and, therefore, result in
contributions $g_{ew}^3\alpha_s\, \delta^{-1}\sim \delta$, i.e. of the
same order as the leading soft contribution. Thus, the subleading hard
part of the self-energy diagram has to be included (but not
necessarily resummed) independent of the renormalization scheme
adopted.

To obtain the scaling of the EW one-loop self-energy insertion we
simply have to replace $\alpha_s$ by $\alpha_{ew}$. Thus, diagram~(2b)
scales as $\delta^{1/2}$ and $\delta^{3/2}$ with hard and soft EW
self-energy insertions respectively.  Therefore, the hard part is not
suppressed with respect to the tree-level amplitude and has to be
resummed. The soft part contributes beyond the accuracy of our
calculation and can be neglected.  Obviously, the resummation of the
hard part of the EW self-energy insertion corresponds to the
resummation indicated in \Eqn{prop_resum}.

For diagram~(2c), the decomposition into hard and soft part is trivial
in that the soft part results in scaleless integrals and therefore
vanishes, whereas the hard part corresponds to the full diagram. This
is not surprising as the loop correction in this diagram is not
affected by the instability of the top quark.

Diagrams~(2d) and (2e) are more interesting and both behave in a
similar way. The soft and hard parts scale as $g_{ew}^3\alpha_s\,
\delta^{-1}\sim \delta$ and contribute at NLO. From an ET point of
view, the hard part of diagram~(2d) contributes to the matching
coefficient of the production operator (or to the decay operator in
the case of diagram~(2e)), whereas the soft part is reproduced by a
loop diagram in the effective theory.

Finally, performing a similar expansion for the box diagram~(2f), it can
be seen that the hard part scales as $\delta^2$ and thus contributes
beyond the accuracy of our calculation. The soft part, however, scales
as $\delta$ and must be included at NLO. This illustrates the
simplifications that can be achieved in the calculation. Rather than
having to compute a full box diagram with several scales, we end up
computing only the soft part, which is much simpler.

\subsection{Real corrections}
\label{sec:real}

For the tree-level and virtual calculations, an ET approach can be used
in a straightforward manner. If one is only interested in the total
cross section, real corrections can also be tackled in this way by
relating the total cross section to the imaginary part of the
forward-scattering amplitude.  However, here we are interested in
computing an arbitrary infrared-safe observable. Hence, we want to
compute the real corrections without explicitly specifying the
observable.

It is clear that, to a certain extent, this is in conflict with an ET
approach. An effective theory relies on making all scales explicit.
Since the definition of the observable itself can introduce additional
scales, it is not possible to follow a strict ET approach.  In this
subsection we describe how we deal with the real corrections, making
sure that we keep all terms to the desired accuracy, for a general
observable. The only assumption we make on the observable is that it
does not introduce another hierarchy of scales. We stress that this
assumption has to be made for any fixed-order calculation because a
large ratio of scales usually introduces large logarithms which have
to be resummed.

\begin{figure}[t!]
\begin{center}
\includegraphics[width=0.9 \linewidth]{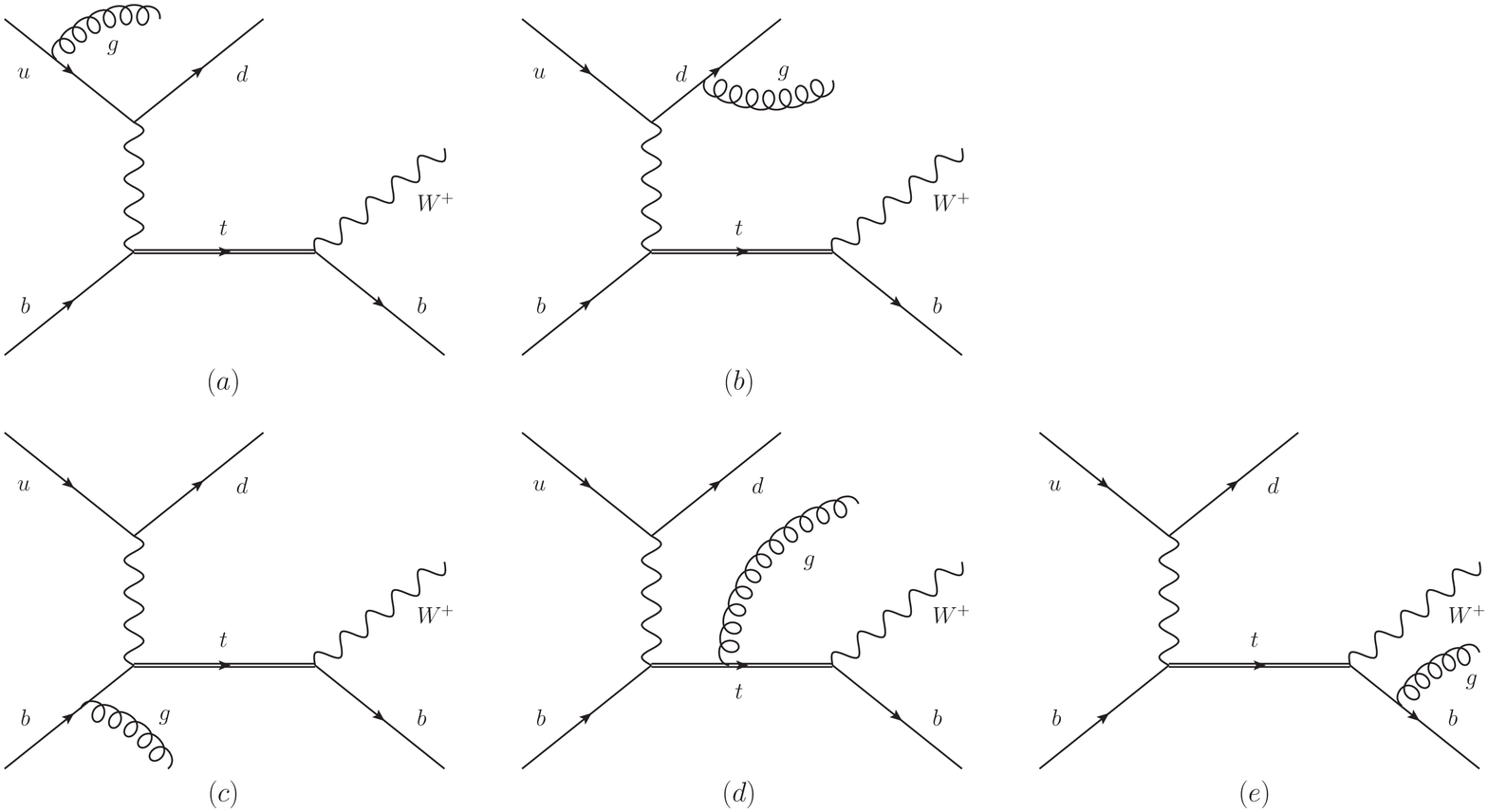}
\end{center}
\ccaption{}{Real QCD corrections to resonant t-channel single-top
  production at leading order in $\Delta_t/m_t$.\label{fig:real}}
\end{figure}

To begin with, we need the real matrix element squared for the process
(\ref{process:ubg}). We show the diagrams containing a top-quark propagator
in Figure~\ref{fig:real}. There are additional diagrams corresponding
to diagrams~(b) and (c) of Figure~\ref{fig:qWbtree} with an additional
gluon radiated off, however, they are suppressed and contribute only
beyond the accuracy we are aiming at.  The real corrections have an
additional factor of $\alpha_s$, which allows us to work at leading order
in $\Delta_t/m_t$. 

In the case of the real corrections it is not, a priori, clear what
the correct expansion parameter is. Diagrams~(a)--(c) have a resonant
top propagator for $(p_W+p_4)^2 \simeq m_t^2$, diagram~(e) has a
resonant top propagator for $(p_W+p_4+p_g)^2 \simeq m_t^2$, and
diagram~(d) is resonant in both kinematic configurations.  Depending
on how the final state partons are combined to jets, both regions can
be relevant and have to be taken into account. This can be achieved by
a slight modification of the usual subtraction method.

In order to isolate infrared singularities, real corrections are
usually computed by subtracting from the real, $n+1$ parton matrix
element squared, $M_{n+1}$, a term $M^{\rm sing}_{n(+1)}$ that
approximates the full matrix element in all singular regions. This
term has essentially $n$-parton kinematics. It is added back and a
partial phase-space integration is performed analytically to recover
the infrared $1/\epsilon$ poles explicitly. The kinematic
configurations associated with $M^{\rm sing}_{n(+1)}$ correspond
either to a gluon being soft or two partons being collinear. Thus, for
$M^{\rm sing}_{n(+1)}$ it is always clear whether the gluon is
combined with the $b$-quark or not and, correspondingly, what the
appropriate expansion parameter is. Therefore, we can expand the term
that is added back and write for the real corrections
\begin{eqnarray}
\int d\Phi_{n+1}  M_{n+1} &=& 
\int d\Phi_{n+1} \left(M_{n+1}- M^{\rm sing}_{n(+1)} \right)
+ \int d\Phi_{n+1}\,  M^{\rm sing }_{n(+1)}
\label{realM}\\
&\simeq& 
\int d\Phi_{n+1} \left( M_{n+1} - M^{\rm sing}_{n(+1)} \right)
+ \int d\Phi_{n+1}\,  M^{\rm sing\ {exp} }_{n(+1)}\, .
\nonumber
\end{eqnarray}
The explicit infrared poles in $M^{\rm sing\ {exp} }_{n(+1)}$ now
match the infrared poles of the virtual corrections. The error
introduced in the second line of \Eqn{realM} is suppressed by a factor
$\alpha_s\, \delta$ relative to the Born term and, therefore, is
beyond the accuracy we are aiming at. It corresponds to the error
introduced in expanding the virtual corrections. Treating the real
corrections in this way, we ensure that the expansion of the real and
virtual corrections in $\Delta_t/m_t$ is performed in a consistent way,
such that the infrared poles always match. We have implemented this
procedure in two independent programs, using the
FKS~\cite{Frixione:1995ms} and dipole subtraction~\cite{Catani:1996vz}
methods. The agreement of the results produced with the two programs
provides us with a useful check of the implementation.

\section{Helicity amplitudes for single-top production}
\label{sec:amplitudes}

In this section we present all the helicity amplitudes necessary for
the computation of the ${\cal O}(\delta)$ and ${\cal O}(\delta^{3/2})$
contributions to the cross section of the process
(\ref{eq:lhc_process}). These consist of the leading resonant
tree-level diagram~(1a) shown in Figure~\ref{fig:qWbtree}, which
scales as $\delta^{1/2}$, and of virtual and real QCD corrections to
this diagram (including the gluon-initiated processes
\Eqns{process:gbu}{process:ugb}), which scale parametrically as
$\delta^{1/2} \alpha_s \sim \delta$. The resummation of finite-width
effects, illustrated in \Eqn{eq:DPA}, also requires the calculation of
one-loop electroweak self-energies, as discussed in
Sections~\ref{sec:virtual} and \ref{sec:ren}. Any other contribution
is suppressed by at least $\delta$ compared to the leading tree-level
process and is beyond the accuracy pursued in this work. The
inclusion of real corrections in our formalism has been discussed in
the previous section, and here we limit ourselves to giving a list of
the relevant amplitudes. For the virtual corrections we will explain,
in some detail, the application of the method of regions to the
computation of the expansion in $\delta$ of loop integrals. Throughout
this section, and the rest of the paper, we adopt the helicity
notation introduced in Ref.~\cite{Xu:1986xb,Mangano:1990by} and make
use of the following abbreviations: $s_{ij} \equiv (p_i+p_j)^2$, $p_t
\equiv p_4+p_5+p_6$, $D_t\equiv p_t^2-m_t^2$ and $\Delta_t \equiv
p_t^2-\mu_t^2$.

\subsection{Tree-level amplitude}

As pointed out in Section~\ref{sec:setup}, of the tree-level
diagrams shown in Figure~\ref{fig:qWbtree} only the resonant one,
diagram~(\ref{fig:qWbtree}a), contributes to the amplitude at
${\cal O}(\delta^{1/2})$. Given its purely electroweak nature, the only
non-vanishing amplitude is the one for the helicity configuration $u_L
b_L \rightarrow d_L b_L e^+_R \nu_L$ and it reads
\begin{equation}
\label{eq:Atree}
g_{ew}^3 A^{(3,0)}_{(-1)} = g_{ew}^4 \sqrt{\frac{\pi}{M_W \Gamma_W}} \,  
\frac{[12]\langle46\rangle\langle3|4+6|5]}{(s_{13}+M_W^2) \Delta_t} \, .
\end{equation}
In \Eqn{eq:Atree}, the extra power of the coupling constant, $g_{ew}$,
and the prefactor $\sqrt{\pi/M_W \Gamma_W}$ arise from the inclusion
of the $W$-boson decay in the improved narrow-width approximation;
$1/((p_W^2-M_W^2)^2+M_W^2 \Gamma_W^2) \sim \pi/(M_W \Gamma_W) \delta
(p_W^2- M_W^2)$. Note, however, that this prefactor does not modify
the scaling of the amplitude since $\Gamma_W \propto g_{ew}^2$.  The
complex pole, $\mu_t^2 \equiv m_t^2-i m_t \Gamma_t$, in the top-quark
propagator, $\Delta_t$, resums leading finite-width effects and can be
related order-by-order in $\alpha_s$, $\alpha_{ew}$ to the
renormalized mass and self-energy of the top quark. This is explicitly
shown below in Section~\ref{sec:ren}.  The amplitude
$A^{(3,0)}_{(-1)}$ is formally gauge independent, up to terms
suppressed by $\delta$ or higher.  These gauge-violating terms, which
would normally be cancelled by the electroweak background diagrams in
Figure~\ref{fig:qWbtree}, are numerically small as long as the
condition $\delta \ll 1$ is satisfied.  For the input parameters and
cuts adopted in Section~\ref{sec:results}, they amount to a correction
to the leading resonant contribution to the cross section of much less
than 1\%. However, when $\delta \gtrsim 1$ the effective-theory
expansion breaks down and these parametrically suppressed
contributions can be numerically important.  In this case it is
necessary to calculate the complete gauge-invariant set of diagrams,
shown in Figure~\ref{fig:qWbtree}.

\subsection{Virtual corrections }
\label{sec:virtual}

The relevant one-loop contributions to the amplitude up to ${\cal O}(\delta)$
are shown in Figure~\ref{fig:1loop}. As mentioned in
Section~\ref{sec:et}, we use the method of regions
\cite{Beneke:1997zp} to compute these loop integrals. This
automatically yields the expansion of the full integral in $\delta$
and has the advantage of simplifying the calculation. In the case of
interest here, the two relevant momentum regions are \emph{soft} ($q_0
\sim \vec{q} \sim \delta$) and \emph{hard} ($q_0 \sim \vec{q} \sim
1$).  Only the hard part of the electroweak self-energy
diagram~(\ref{fig:1loop}b) contributes to the amplitude at order
$\delta$, and this is automatically included in the cross section
through the resummation of finite-width effects in the top-quark
propagator, as explained in Section~\ref{sec:ren}. Therefore, we
focus here on the remaining QCD diagrams, (\ref{fig:1loop}a) and
(\ref{fig:1loop}c)--(\ref{fig:1loop}f).

To illustrate how the expansion by regions works we will now
explicitly show how it is applied to the case of
diagram~(\ref{fig:1loop}d), starting with the computation of the soft
part.  The full expression for the three-point loop integral is
\begin{equation} \label{eq:region_ex}
A^{(3,2)}_{(-1),d} = \alpha_s C_F \int \frac{d^d q}{(2 \pi)^d} 
\frac{...(\dirac{p}_t-m_t) \gamma^\mu 
(\dirac{p}_t-\dirac{q}-m_t)...(\dirac{p}_2-\dirac{q}) 
 \gamma_\mu u(p_2)}{q^2  (q^2-2 p_t \cdot q+D_t)(q^2-2 p_2 \cdot q)} \, ,
\end{equation} 
where the ellipses denote constants and spinor quantities which do not
depend on the loop momentum, $q$. Let us first consider the expansion
of the three propagators appearing in the denominator. The gluon
propagator has an homogeneous scaling, since all the components of $q$
are of order $\delta$, and is not expanded. In the light-fermion
propagator, $q^2-2 p_2 \cdot q$, the quadratic term ($\sim \delta^2$) is
suppressed compared to the linear term ($\sim \delta$) and can be
dropped. Therefore, $q^2-2 p_2 \cdot q \rightarrow -2 p_2 \cdot q$. In the
top-quark propagator, the term quadratic in $q$ can again be neglected,
whereas $p_2 \cdot q$ and $D_t$ both scale as $\delta$. Furthermore,
given that $D_t-2 p_t \cdot q \sim \delta$ in the soft region,
finite-width effects must be resummed to all orders in the top-quark
propagator, leading to
\begin{equation}
\frac{1}{q^2-2 p_t \cdot q+D_t} \rightarrow  
\frac{1}{\Delta_t-2 p_t \cdot q} \, .
\end{equation} 
In the numerator of \Eqn{eq:region_ex}, the loop momentum, $q$, is always
parametrically smaller than the external momenta, $p_i$, and can be
neglected at leading order in $\delta$. From the properties of the
Dirac algebra and of the spinors, $u(p)$, it also follows that
\begin{equation}
\dirac{p}_2 \gamma_\mu u(p_2) = 2 p_{2,\mu} u(p_2) \, ,
\end{equation}
\begin{equation}
(\dirac{p}_t-m_t) \gamma^\mu (\dirac{p}_t-m_t)=2 p_t^\mu (\dirac{p}_t-m_t)
-D_t \gamma^\mu \sim 2 p_t^\mu (\dirac{p}_t-m_t) \, ,
\end{equation}
where we have used $D_t \sim \delta$ to drop the second term.
Thus, the leading soft contribution to the loop integral
(\ref{eq:region_ex}) is
\begin{equation} \label{eq:region_ex_S}
A^{(3,2)S}_{(-1),d} = \alpha_s C_F (4 p_2 \cdot p_t) 
\int \frac{d^d q}{(2 \pi)^d} 
\frac{...(\dirac{p}_t-m_t) ... u(p_2)}
{q^2  (\Delta_t-2 p_t \cdot q)  (-2 p_2 \cdot q)} \, .
\end{equation}
Note that the leading soft part in the expansion by regions is
equivalent to the well known eikonal approximation,
\begin{equation}\label{eq:eikon}
\frac{(\dirac{p}_2-\dirac{q}) \gamma^\mu}{(p_2-q)^2} 
\rightarrow \frac{p_2^\mu}{(-p_2 \cdot q)} \, .
\end{equation}
Eq. (\ref{eq:region_ex_S}) is much simpler than the original integral
(\ref{eq:region_ex}) and can be easily computed with standard
techniques. The explicit result is given in Eq. (\ref{eq:soft}).  The
parametric scaling of the correction (\ref{eq:region_ex_S}) can
actually be checked before the explicit calculation of the
integral. Given that a soft momentum scales as $\delta$; the gluon
propagator scales as $\delta^{-2}$, the light-quark and top-quark
propagator as $\delta^{-1}$ and the infinitesimal volume element $d^4
q$ as $\delta^4$. Thus, \Eqn{eq:soft} is suppressed compared
to the leading amplitude by $\alpha_s \times \delta^4 \times
\delta^{-2} \delta^{-1} \delta^{-1} \sim \delta^{1/2}$, as expected.

The remaining QCD diagrams in Figure~\ref{fig:1loop} can be expanded
in the soft region analogously to diagram~(\ref{fig:1loop}d). As a
consequence of the simple structure of the quark-gluon vertex in the
soft limit, \Eqn{eq:eikon}, the soft corrections factorize in
terms of scalar functions and the leading tree-level amplitude,
$A^{(3,0)}_{(-1)}$.  The contributions of the different diagrams in
Figure~\ref{fig:1loop} read as follows ($\tilde{\mu}^{2 \epsilon} \equiv
e^{\epsilon \gamma_E} \mu^{2 \epsilon}/(4 \pi)^\epsilon$ with $\mu$
being the renormalization scale):
{\allowdisplaybreaks
\begin{eqnarray}\label{eq:soft}
A^{(3,2)S}_{(-1),a} &=& \left[-16 \pi i \alpha_s C_F
\tilde{\mu}^{2 \epsilon} \frac{m_t^2}{\Delta_t}  \int
\frac{d^d q}{(2 \pi)^d} \frac{1}{q^2} 
\frac{1}{\Delta_t-2 p_t \cdot q} \right] A^{(3,0)}_{(-1)}\nonumber\\
&=& \frac{\alpha_s C_F}{2 \pi} \left[\frac{1}{\epsilon}+2 \right]
\left(-\frac{\Delta_t}{\mu m_t}\right)^{-2 \epsilon} A^{(3,0)}_{(-1)} \nonumber\\
&& \nonumber\\
A^{(3,2)S}_{(-1),c} &=&
\left[-16 \pi i \alpha_s C_F \tilde{\mu}^{2 \epsilon} (p_1 \cdot p_3) \int
\frac{d^d q}{(2 \pi)^d} \frac{1}{q^2} \frac{1}{(-2 p_1 \cdot q)}
\frac{1}{(-2 p_3 \cdot q)}\right] A^{(3,0)}_{(-1)}\nonumber\\
&=&0 \nonumber\\
&& \nonumber\\
A^{(3,2)S}_{(-1),d} &=& \left[-16 \pi i \alpha_s C_F
\tilde{\mu}^{2 \epsilon} (p_2 \cdot p_t)  \int
\frac{d^d q}{(2 \pi)^d} \frac{1}{q^2} \frac{1}{(-2 p_2 \cdot q)}
\frac{1}{\Delta_t-2 p_t \cdot q}\right] A^{(3,0)}_{(-1)}\nonumber\\
&=& \frac{\alpha_s C_F}{2 \pi} \left[\frac{1}{2 \epsilon^2}
+\frac{5}{24} \pi^2 \right]
\left(-\frac{\Delta_t}{\mu m_t}\right)^{-2 \epsilon} 
A^{(3,0)}_{(-1)}\nonumber\\
&& \nonumber\\
A^{(3,2)S}_{(-1),e} &=& \left[-16 \pi i \alpha_s C_F
\tilde{\mu}^{2 \epsilon} (p_4 \cdot p_t)  \int
\frac{d^d q}{(2 \pi)^d} \frac{1}{q^2} \frac{1}{(-2 p_4 \cdot q)}
\frac{1}{\Delta_t-2 p_t \cdot q}\right] A^{(3,0)}_{(-1)}\nonumber\\
&=& \frac{\alpha_s C_F}{2 \pi} \left[\frac{1}{2 \epsilon^2}
+\frac{5}{24} \pi^2 \right]\left(-\frac{\Delta_t}{\mu m_t}\right)^{-2 \epsilon} A^{(3,0)}_{(-1)}
\nonumber\\
&& \nonumber\\
A^{(3,2)S}_{(-1),f} &=& \left[-16 \pi i \alpha_s C_F
\tilde{\mu}^{2 \epsilon} (p_2 \cdot p_4)  \int
\frac{d^d q}{(2 \pi)^d} \frac{1}{q^2} \frac{1}{(-2 p_2 \cdot q)}
\frac{1}{(-2 p_4 \cdot q)}
\frac{\Delta_t}{\Delta_t-2 p_t \cdot q}\right] A^{(3,0)}_{(-1)}\nonumber\\
&=& \frac{\alpha_s C_F}{2 \pi} \left[-\frac{1}{\epsilon^2} -\frac{1}{\epsilon}
\ln \left( \frac{(s_{2t}-m_t^2) (s_{4t}-m_t^2)}{m_t^2 s_{24}}\right)
\right.\nonumber\\
&&\left.+\text{Li}_2 \left(1
-\frac{(s_{2t}-m_t^2) (s_{4t}-m_t^2)}{m_t^2 s_{24}} \right)
-\frac{5}{12} \pi^2 \right]
\left(-\frac{\Delta_t}{\mu m_t}\right)^{-2 \epsilon}  A^{(3,0)}_{(-1)}\, .
\end{eqnarray} }
In the soft limit, diagram~(2c) reduces to a scaleless integral that vanishes
in dimensional regularization. This is a consequence of the fact that the 
upper quark line of the tree-level diagram (\ref{fig:qWbtree}a) does not
carry any information about the off-shellness of the intermediate
top-quark propagator, and is thus not sensitive to the soft scale, $m_t
\delta$. The total soft correction is 
\begin{eqnarray}
 A^{(3,2)S}_{(-1)}  &=&  \sum_i A^{(3,2)S}_{(-1),i} =
\delta V^S\, A^{(3,0)}_{(-1)}  \, ,
\nonumber
\\
\delta V^S  &=&\frac{\alpha_s C_F}{2 \pi} \left(
-\frac{\Delta_t}{\mu m_t}\right)^{-2 \epsilon}
\left[\frac{1}{\epsilon} \left(1-\ln \left(
\frac{(s_{2t}-m_t^2) (s_{4t}-m_t^2)}{m_t^2 s_{24}}\right) \right)
\right.
\nonumber\\
&& \left.\qquad +\ 2+\text{Li}_2 \left(1
-\frac{(s_{2t}-m_t^2) (s_{4t}-m_t^2)}{m_t^2 s_{24}}\right)\right] \, .
\label{softall}
\end{eqnarray}
In the hard region, the loop momentum scales as $q \sim 1$ and
cannot be neglected in the numerator or denominator of
\Eqn{eq:region_ex}. In this case, however, $p_t^2-m_t^2 \sim \delta$
is much smaller than any other invariant and one can expand the
integrand of \Eqn{eq:region_ex} in $D_t$. The leading hard
contribution for diagram~(\ref{fig:1loop}d) then reads
\begin{equation} \label{eq:region_ex_H}
A^{(3,2)H}_{(-1),d} = \alpha_s C_F \int \frac{d^d q}{(2 \pi)^d} 
\frac{...(\dirac{p}_t-m_t) \gamma^\mu (\dirac{p}_t-\dirac{q}-m_t)...
(\dirac{p}_2-\dirac{q}) \gamma_\mu u(p_2)}
{q^2  (q^2-2 p_t \cdot q)  (q^2-2 p_2 \cdot q)} \, . 
\end{equation}
As for the soft region, the expansion in $\delta$ leads to a
significant simplification of the integrand.  Note that, in the hard
region, there is no resummation of self-energy insertions, since by
assumption $q^2-2 p_t \cdot q \sim m_t^2$. Furthermore, at leading
order in $\delta$, the hard part of (\ref{eq:region_ex}) coincides
with the one-loop vertex correction to the on-shell single-top
production process $u(p_1) b(p_2) \rightarrow d(p_3) t(p_t)$, with
$p_t^2=m_t^2$.  In the effective-theory language this is equivalent to
the statement that the hard matching coefficients are obtained from
the computation of fixed-order on-shell matrix elements. Strictly
speaking, in this context the term ``on-shell'' would imply $p_t^2 =
\mu_t^2$. However, this condition must be satisfied only
order-by-order in $\alpha^2_s \sim \alpha_{ew} \sim \delta$, and
for an ${\cal O}(\delta)$ calculation we can simply set $p_t^2=m_t^2$
in the leading hard contributions.  Using the counting scheme
(\ref{eq:scaling}) we can again determine the scaling behaviour of
\Eqn{eq:region_ex_H}, which is, in this case, $\alpha_s \times 1\times 1
\sim \delta^{1/2}$, since all momenta scale as $\sim 1$. This is
confirmed by the explicit result given below in \Eqn{eq:hard}.

Applying the hard-region expansion to
diagrams~(\ref{fig:1loop}c)--(\ref{fig:1loop}f) we obtain:
\begin{eqnarray}
 A^{(3,2)H}_{(-1),c} &=&  \delta V_{13}^H A^{(3,0)}_{(-1)} \nonumber\\
 A^{(3,2)H}_{(-1),d} &=&  \delta V_{2t}^H A^{(3,0)}_{(-1)}
+\frac{g_{ew}}{2} \sqrt{\frac{\pi}{M_W\Gamma_W}} \,  
\frac{[52]\langle46\rangle\langle3|2|1]}{(s_{13}+M_W^2) \Delta_t} 
\frac{\alpha_s C_F}{2 \pi}
\frac{m_t^2}{2 m_t^2-s_{2t}} \ln\left(\frac{s_{2t}-m_t^2}{m_t^2}\right)  
\nonumber\\
 A^{(3,2)H}_{(-1),e} &=&  \delta V_{4t}^H A^{(3,0)}_{(-1)}
+\frac{g_{ew}}{2} \sqrt{\frac{\pi}{M_W \Gamma_W}} \,  
\frac{[12]\langle43\rangle\langle6|4|5]}{(s_{13}+M_W^2) \Delta_t} 
\frac{\alpha_s C_F}{2 \pi}
\frac{m_t^2}{2 m_t^2-s_{4t}} \ln\left(\frac{s_{4t}-m_t^2}{m_t^2}\right)  
\nonumber\\
A^{(3,2)H}_{(-1),f} &=&  0 
\end{eqnarray}
where the three scalar function $\delta V_{13}^H$, $\delta V_{2t}^H$,
$\delta V_{4t}^H$ are given by:
\begin{eqnarray}\label{eq:hard}
\delta V^H_{13} &=& \frac{\alpha_s C_F}{2 \pi}
\left[-\frac{1}{\epsilon^2}+\frac{1}{\epsilon}
\left( \ln \left(\frac{s_{13}}{\mu^2}\right)-\frac{3}{2}\right)
-\frac{1}{2} \ln^2 \left(\frac{s_{13}}{\mu^2}\right)
+\frac{3}{2} \ln \left(\frac{s_{13}}{\mu^2}\right)-4
+\frac{x_{sc}}{2}+\frac{\pi^2}{12}\right]\nonumber\\
&& \nonumber\\
\delta V^H_{2t} &=& \frac{\alpha_s C_F}{2 \pi}
\left[-\frac{1}{2 \epsilon^2}+\frac{1}{\epsilon}
\left( \ln \left(\frac{s_{2t}-m_t^2}{m_t \mu}\right)
-\frac{1}{2}\right)+
\text{Li}_2 \left(1-\frac{m_t^2}{s_{2t}-m_t^2}\right)-2+
\frac{x_{sc}}{2}-\frac{\pi^2}{24}\right.\nonumber\\
&&-\frac{1}{2} \ln^2 \left(\frac{s_{2t}-m_t^2}{m_t \mu}\right)+
\frac{1}{8} \ln^2 \left(\frac{m_t^2}{\mu^2}\right)+
\frac{s_{2t}-m_t^2}{4 (2 m_t^2-s_{2t})}
\ln \left(\frac{m_t^2}{\mu^2}\right)\nonumber\\
&& +\left. \ln \left(\frac{s_{2t}-m_t^2}{m_t \mu}\right)
\left(1-\frac{s_{2t}-m_t^2}{2(2 m_t^2-s_{2t})}-
\frac{1}{2}\ln \left(\frac{m_t^2}{\mu^2}\right)
\right)\right]\nonumber\\
&& \nonumber\\
\delta V^H_{4t} &=& \delta V^H_{2t} \Big |_{s_{2t} \rightarrow s_{4t}} \, .
\end{eqnarray}
The expressions in \Eqn{eq:hard}, once renormalized as explained in
the next section, agree with the results available in the literature
(see e.g. Eq.~(19) of Ref.~\cite{Harris:2002md} and Eq.~(9) of
Ref.~\cite{Campbell:2004ch}). This is expected, since the leading hard
contributions must coincide with the corresponding corrections to the
on-shell top production process.  The value of the parameter $x_{sc}$
in \Eqn{eq:hard} depends on the particular version of dimensional
regularization used for the calculation (see
e.g. Ref.~\cite{Signer:2008va}), and is given by $x_{sc}=0$ in the 't
Hooft-Veltman scheme (HV) and $x_{sc}=1$ in the four-dimensional
helicity scheme (FDH).  As anticipated in Section~\ref{sec:et}, the
${\cal O}(\alpha_s \delta^{1/2})$ hard contribution of
diagram~(\ref{fig:1loop}f) vanishes since the intermediate top-quark
propagator is off-shell by an amount $\sim m_t^2$. This leads to a
further suppression in $\delta$ compared to the other one-loop QCD
diagrams shown in Figure~\ref{fig:1loop}.  In a strict
effective-theory approach, the contribution of
diagram~(\ref{fig:1loop}f) would be encoded in an effective
higher-dimensional $bW^+\rightarrow bW^+$ vertex.
 
The expansion of the self-energy diagram~(\ref{fig:1loop}a) in the
hard region presents some subtleties.  Let us consider the leading
(fixed-order) top-quark propagator, $i (\dirac{p}_t+m_t)/D_t$, and the
QCD self-energy correction to the propagator,
\begin{equation}\label{eq:selfenergy}
\frac{i (\dirac{p}_t+m_t)}{D_t}\left[-4 \pi \alpha_s C_F
\tilde{\mu}^{2 \epsilon} \int \frac{d^d q}{(2 \pi)^d}
\frac{\gamma^\alpha (\dirac{p}_t-\dirac{q}+m_t) \gamma_\alpha}
{q^2 (q^2 -2 p_t \cdot q+D_t)}
\right]\frac{i (\dirac{p}_t+m_t)}{D_t} \, .
\end{equation}
By applying our counting to \Eqn{eq:selfenergy}, one expects the
leading hard contribution of the QCD self-energy correction to scale
as $\alpha_s \delta^{-2} \sim \delta^{-3/2}$, i.e. it is enhanced by
$\delta^{-1/2}$ with respect to the leading-order propagator, which
scales as $\delta^{-1}$. Hence, in this case we have to push the
expansion by regions one order further to include all the terms
suppressed by $\delta^{1/2}$ compared to the leading tree-level
contribution. Thus, the expansion of \Eqn{eq:selfenergy} in $D_t$, up to
the relevant order, is
\begin{equation} \label{eq:self_expansion}
\frac{\alpha_s C_F}{2 \pi} 
\left[\frac{3}{2 \epsilon}+2+\frac{x_{sc}}{2}\right]
\left(\frac{m_t^2}{\mu^2}\right)^{-\epsilon} 
\left[\frac{2 i m_t^2 (\dirac{p}_t+m_t)}{D_t^2}+
\frac{i m_t}{D_t}-\frac{i(\dirac{p}_t+m_t)}{D_t}\right] \, .
\end{equation}   
As anticipated, the first term in \Eqn{eq:self_expansion} is
``superleading'' and in a generic renormalization scheme has to be
resummed in the top-quark propagator. However, in the on-shell scheme
adopted in this paper, the superleading bit is cancelled exactly by the
top-quark mass counterterm, as shown in Section~\ref{sec:ren}.

As pointed out at the beginning of this section, only the hard part of
the electroweak self-energy diagram (\ref{fig:1loop}b) is relevant to
our calculation since a soft contribution would scale as
$\delta^{1/2} \alpha_{ew} \sim \delta^{3/2}$. In this case, the
dominant hard correction to the propagator scales as $\alpha_{ew}
\delta^{-2} \sim \delta^{-1}$ and is resummed, while higher-order
terms are beyond the target accuracy pursued here and can be
neglected.  In principle, diagrams analogous to (\ref{fig:1loop}b), but
with a photon or a $Z$-boson in the loop, should also be resummed. 
As we will see in the next section, in the on-shell scheme only
the imaginary part of the one-loop two-point function is effectively
resummed in the propagator, and diagram~(\ref{fig:1loop}b) is therefore
sufficient.
 
\subsection{Renormalization and resummation of finite-width effects}
\label{sec:ren}

We now discuss the renormalization of the one-loop amplitudes computed
in Section~\ref{sec:virtual}. In this paper we adopt the on-shell
scheme (see e.g. Refs.~\cite{Denner:1991kt, Bohm:2001yx}). In
this scheme, the wave-function counterterms for external massless
particles vanish identically, i.e. $\delta Z_i=0$ for
$i=1,...,6$. Furthermore, the electroweak coupling, $g_{ew}$, is not
renormalized at ${\cal O}(\alpha_s)$. Thus, the only counterterm
relevant for our calculation is the top-quark mass counterterm,
$\delta m_t$. This induces a renormalization of the top-quark
propagator, which is
\begin{equation}\label{eq:ren_prop} 
\frac{i (\dirac{p}_t+m_t)}{D_t} (-i \delta m_t) 
\frac{i (\dirac{p}_t+m_t)}{D_t} = \frac{\delta m_t}{m_t}
\left[\frac{2 i m_t^2 (\dirac{p}_t+m_t)}{D_t^2}+\frac{i m_t}{D_t}\right] \, ,
\end{equation}
with the ${\cal O}(\alpha_s)$ mass counterterm given by 
\begin{equation}
\frac{\delta m_t}{m_t} = 
-\frac{\alpha_s C_F}{2 \pi} 
\left[\frac{3}{2 \epsilon}+2+\frac{x_{sc}}{2}\right]
\left(\frac{m_t^2}{\mu^2}\right)^{-\epsilon} \, .
\label{deltam}
\end{equation}
Comparing \Eqns{eq:self_expansion}{eq:ren_prop}, it is clear
that the superleading terms (and part of the subleading ones) arising
from the self-energy correction are cancelled by the
counterterm. Rewriting \Eqns{eq:self_expansion}{eq:ren_prop} as
contributions to the amplitude, one obtains
\begin{equation}
A_{(-1),a}^{(3,2)H}+A_{(-1)}^{(3,2),\text{ren}}=
\frac{\alpha_s C_F}{2 \pi} \left[-\frac{3}{2 \epsilon}
-2 -\frac{x_{sc}}{2}\right] 
\left(\frac{m_t^2}{\mu^2}\right)^{-\epsilon} A_{(-1)}^{(3,0)} \, .
\end{equation}

In an arbitrary renormalization scheme, which in the following we will
denote by $r$, there is, in general, no exact cancellation of the
superleading contributions in \Eqns{eq:self_expansion}{eq:ren_prop},
and these terms are resummed inside the complex pole of the top-quark
propagator, $\mu_t^2 \equiv m_t^2-i m_t \Gamma_t$.  Consider the
renormalized top-quark two-point function,
\begin{equation} \label{eq:inv_prop}
\overline{\Gamma}_{t}(\dirac{p})\equiv 
\dirac{p}-m_{t,r}+\overline{\Sigma}_{t,r}(\dirac{p}) \, ,
\end{equation}
where $m_{t,r}$ is the renormalized mass in the generic scheme $r$
and $\overline{\Sigma}_{t,r}$ represents the renormalized top-quark
self-energy.  The complex pole, $\mu_t$, is defined as the solution for
$\dirac{p}$ at which \Eqn{eq:inv_pole_rew} vanishes (see
e.g. Ref.~\cite{Jegerlehner:2003py}).  Since $\Delta_t \sim \delta$,
\Eqn{eq:inv_prop} must include all contributions to
$\overline{\Sigma}_{t,r}$ up to ${\cal O}(\delta)$, i.e. the one-loop
QCD ($\sim \alpha_s \sim \delta^{1/2}$) and electroweak ($\sim
\alpha_{ew} \sim \delta$) top-quark self-energies, and the two-loop
QCD contribution ($\sim \alpha_s^2 \sim \delta$). Other corrections to
$\overline{\Sigma}_{t,r}$ can be included perturbatively and need not
to be resummed.

In the following we will adopt the notation of
Ref.~\cite{Denner:1991kt} and parametrize the top-quark self-energy as
\begin{equation}\label{eq:self}
\overline{\Sigma}_{t,r}(\dirac{p}) = 
\overline{\Sigma}_{t,r}^L(p^2) \dirac{p} P_L
+ \overline{\Sigma}_{t,r}^R(p^2) \dirac{p} P_R 
+\overline{\Sigma}_{t,r}^S(p^2) m_{t,r} \, . 
\end{equation} 
The renormalized quantities in \Eqn{eq:self} are related to the
unrenormalized ones by:
\begin{eqnarray}
\overline{\Sigma}_{t,r}^L(p^2) &=& \Sigma_t^L(p^2)+\delta Z^L_{t,r} \nonumber\\
\overline{\Sigma}_{t,r}^R(p^2) &=& \Sigma_t^R(p^2)+\delta Z^R_{t,r} \nonumber\\
\overline{\Sigma}_{t,r}^S(p^2) &=& 
\Sigma_t^S(p^2)-\frac{\delta m_{t,r}}{m_{t,r}}-\frac{\delta Z^L_{t,r}+
\delta Z^R_{t,r}}{2} \, ,
\end{eqnarray}
where $\delta m_{t,r}$ and $\delta Z^{L/R}_{t,r}$ denote the mass and
wave-function counterterms, in the scheme $r$, respectively.
\Eqn{eq:inv_prop} can be rewritten in a form more suitable to the
extraction of the pole by inserting identity matrices, in the form
${\bf I}=(\dirac{p}-m_{t,r}) (\dirac{p}+m_{t,r})/(p^2-m_{t,r}^2)$, at
both sides of $\overline{\Sigma}_{t,r}$ and expanding the quantity
$(\dirac{p}+m_{t,r}) \overline{\Sigma}_{t,r} (\dirac{p}+m_{t,r})$
around $p^2 =m_{t,r}^2$. This leads to
\begin{eqnarray} \label{eq:inv_pole_rew}
&&\overline{\Gamma}_t(\dirac{p})=\dirac{p}-m_{t,r} 
+m_{t,r}^2 \overline{\Omega}_{t,r}(p^2)
\frac{\dirac{p}-m_{t,r}}{p^2-m_{t,r}^2} \nonumber\\
&&\hspace{2 cm}+m_{t,r}^2 \left[ 2 \overline{\Sigma}_{t,r}^L(p^2)
+2 \overline{\Sigma}_{t,r}^R(p^2)+\overline{\Omega}_{t,r}(p^2)\right]
\frac{(\dirac{p}-m_{t,r})^3}{(p^2-m_{t,r}^2)^2} +... \, ,
\end{eqnarray} 
where the ellipsis indicates terms that are suppressed beyond ${\cal
  O(\delta)}$ and, thus, are irrelevant to the calculation presented
here. For later convenience, in \Eqn{eq:inv_pole_rew} we have
introduced the function $\overline{\Omega}_{t,r} =
\overline{\Sigma}_{t,r}^L + \overline{\Sigma}_{t,r}^R+2
\overline{\Sigma}_{t,r}^S$.
 
The complex mass, $\mu_t$, is obtained by solving the equation
$\overline{\Gamma}_t(\mu_t)=0$, and reads
\begin{equation} \label{eq:mut}
\mu_t = m_{t,r} -\frac{m_{t,r}}{2} \overline{\Omega}_{t,r}(m_{t,r}^2)
+\frac{m_{t,r}}{4} \overline{\Omega}_{t,r}(m_{t,r}^2)
\left[ \overline{\Sigma}_{t,r}^L(m_{t,r}^2)
+\overline{\Sigma}_{t,r}^R(m_{t,r}^2)+2 m_{t,r}^2 
\frac{\partial \overline{\Omega}_{t,r}}{\partial p^2} (m_{t,r}^2) \right] \, .
\end{equation}
\Eqn{eq:mut} can be further expanded in $\alpha_s$ and
$\alpha_{ew}$. To this end, we introduce a decomposition of the
self-energies analogous  to \Eqn{treeA}
\begin{eqnarray} \label{eq:Sigma_coup}
\overline{\Sigma}^S_{t,r} &=& \alpha_s
\overline{\Sigma}^{S,(0,2)}_{t,r}+
\alpha_s^2 \overline{\Sigma}^{S,(0,4)}_{t,r}+\alpha_{ew}
\overline{\Sigma}^{S,(2,0)}_{t,r} \nonumber\\
\overline{\Sigma}^{L/R}_{t,r} &=& \alpha_s
\overline{\Sigma}^{V,(0,2)}_{t,r}+
\alpha_s^2 \overline{\Sigma}^{V,(0,4)}_{t,r}+\alpha_{ew}
\overline{\Sigma}^{L/R,(2,0)}_{t,r} \, .
\end{eqnarray}
Note that the pure QCD part of the self-energy does not have an axial
component, and
$\overline{\Sigma}^{L,(0,n)}_{t,r}=\overline{\Sigma}^{R,(0,n)}_{t,r}
\equiv \overline{\Sigma}^{V,(0,n)}_{t,r}$ for $n=2,4$. Using
\Eqns{eq:mut}{eq:Sigma_coup} one can easily obtain the following
results for the pole mass and width:
\begin{eqnarray} \label{eq:mtGammat}
m_t \equiv \text{Re}[\mu_t] &=& m_{t,r}- \alpha_s m_{t,r}
\left(\overline{\Sigma}^{V,(0,2)}_{t,r}
+\overline{\Sigma}^{S,(0,2)}_{t,r}\right)
-\alpha_s^2 m_{t,r}  \left(\overline{\Sigma}^{V,(0,4)}_{t,r}
+\overline{\Sigma}^{S,(0,4)}_{t,r}\right)\nonumber\\
&&+\alpha_s^2 m_{t,r} \left(\overline{\Sigma}^{V,(0,2)}_{t,r}
+\overline{\Sigma}^{S,(0,2)}_{t,r}\right)
\left[\overline{\Sigma}^{V,(0,2)}_{t,r}
+2 m_{t,r}^2 \left(\frac{\partial
\overline{\Sigma}_{t,r}^{V,(0,2)}}{\partial p^2}+
\frac{\partial \overline{\Sigma}_{t,r}^{S,(0,2)}}{\partial
p^2}\right)\right]\nonumber\\
&&-\alpha_{ew} \frac{m_{t,r}}{2}
\left(\text{Re}\left[\overline{\Sigma}^{L,(2,0)}_{t,r}\right]+
\overline{\Sigma}_{t,r}
+2 \overline{\Sigma}^{S,(2,0)}_{t,r}\right) \nonumber\\
\Gamma_t \equiv -2 \text{Im}[\mu_t] &=& \alpha_{ew} m_{t,r}
\text{Im} \left[\overline{\Sigma}^{L,(2,0)}_{t,r}\right] \, ,
\end{eqnarray}
where we have used the information that only
$\overline{\Sigma}^{L,(2,0)}_{t,r}$ has a non-vanishing imaginary
part, and all the functions are evaluated at $p^2=m_{t,r}^2$.

In the on-shell scheme (os), \Eqn{eq:mtGammat} assumes a
particularly simple structure:
\begin{eqnarray} \label{eq:onshellmass}
m_t &=& m_{t,\text{os}} \nonumber\\
\Gamma_t &=& \alpha_{ew} m_{t,\text{os}} \text{Im}
[\Sigma^{L,(2,0)}_t(m_{t,\text{os}}^2)] \, .
\end{eqnarray}
The result (\ref{eq:onshellmass}) follows from the particular form of
the mass and field counterterms in this scheme \cite{Denner:1991kt}:
\begin{eqnarray}
\delta m_{t,\text{os}} &=& \frac{m_{t,\text{os}}}{2} 
\text{Re}\left[\Sigma_t^L+\Sigma_t^R+2 \Sigma_t^S \right] 
\nonumber\\
\delta Z_{t,\text{os}}^{L,R} &=& 
-\text{Re} \left[\Sigma_t^{L/R} +m_{t,\text{os}}^2 
\left(\frac{\partial \Sigma_t^L}{\partial p^2}
+\frac{\partial \Sigma_t^R}{\partial p^2}
+2 \frac{\partial \Sigma_t^S}{\partial p^2}\right)\right] \, .
\end{eqnarray}
Thus, at the level of accuracy we are interested in, the pole mass,
$m_t$, can be identified with the on-shell mass, $m_{t,\text{os}}$
\footnote{It is well known that the relation (\ref{eq:onshellmass}) is
  corrected by terms of order $\alpha_{ew}^2$, which are, however,
  beyond the accuracy pursued here.}. Furthermore, as anticipated at
the end of Section \ref{sec:virtual}, only the imaginary part of the
one-loop electroweak self-energy is effectively resummed in the
propagator. In particular, in \Eqn{eq:onshellmass} the dependence from
the one-loop (and two-loop) QCD contributions, which are responsible
for the superleading terms in \Eqns{eq:self_expansion}{eq:ren_prop},
cancels completely. The imaginary part of the complex pole is given,
as expected, by the on-shell top decay width
\begin{equation}
\Gamma_t = \frac{G_F}{8 \pi \sqrt{2}} m_t^3 
\left(1-\frac{M_W^2}{m_t^2}\right)^2 
\left(1+2 \frac{M_W^2}{m_t^2}\right) \, .
\end{equation}    

The residue of the propagator at the complex pole, $\delta R_t$,
which enters \Eqn{eq:DPA}, can be easily extracted from
\Eqn{eq:inv_pole_rew} and expanded in $\alpha_s$ and $\alpha_{ew}$ up
to the required accuracy. Here we only mention the fact that in a
generic scheme $\delta R_{t,r} \sim \alpha_s+...$, and the term
$\delta R_{t,r}\, A^{(3,0)}_{(-1)}$ is parametrically of the same order
as the one-loop QCD corrections computed in Section~\ref{sec:virtual}.
Thus, it also contributes to the amplitude at order $\delta$.  However,
in the on-shell scheme, the expansion of $\delta R_{t,\text{os}}$
starts at order $\alpha_{ew}$ and contributes a correction to the
amplitude which is beyond the accuracy pursued here.

We would like to conclude this section with a remark on how a ``bad''
choice of renormalization scheme can lead to a breakdown of the
effective-theory counting scheme, \Eqn{eq:scaling}. Throughout this
paper we assumed the scalings $D_t \equiv p_t^2-m_{t,r}^2 \sim \delta$
and $\Delta_t \equiv p_t^2-\mu_t^2 \sim \delta$ for the bare and
resummed top-quark propagator. This is consistent in the on-shell
scheme as $\mu_t^2 -m_{t,\text{os}}^2 = \mu_t^2-m_t^2 \sim
\delta$. However, in a generic renormalization scheme the two
conditions are incompatible since $\mu_t^2-m_{t,r}^2 \sim \alpha_s
\sim \sqrt{\delta}$.  This is a problem well known in the context of
applying effective-theory methods to quarkonium physics.  Thus, while
in principle one could choose an arbitrary renormalization scheme
(e.g. $\overline{\text{MS}}$), in practice this can lead to
complications and loss of transparency in the expansion in $\delta$.
In other words, the effective-theory approach adopted here naturally
identifies a class of ``good'' renormalization schemes defined by the
condition $m_{t,r}-m_t \sim \delta$, of which the on-shell scheme
represents a particular example.

\subsection{Real corrections}

The last class of ${\cal O}(\delta^{1/2} \alpha_s)$ contributions to the
amplitude we have to consider are represented by real gluonic
corrections to the tree-level process (\ref{eq:lhc_process}), i.e. $u
(p_1) b(p_2) \rightarrow d(p_3) b(p_4) l^+(p_5) \nu_l(p_6)
g(p_7)$. The relevant Feynman diagrams are shown in
Figure~\ref{fig:real}. The (leading) real-emission amplitude can be
written as
\begin{equation}
{\cal A}^{\text{real}}_{(-1)}(g_7^{\pm}) = 
\delta_{24} T^{a_7}_{31} g_{ew}^3 g_s A^{(3,1)}_{[31]}(g_7^{\pm})
+\delta_{13} T^{a_7}_{42} g_{ew}^3 g_s A^{(3,1)}_{[42]}(g_7^{\pm}) \, ,
\end{equation} 
where $g_7^{\pm}$ denotes the two possible helicity states of the
emitted gluon, and $A^{(3,1)}_{[31]}$ and $A^{(3,1)}_{[42]}$ represent
the contribution of diagrams with a gluon attached to the upper or
lower fermion line, respectively.  The helicity amplitudes for the
upper-line emission read
\begin{eqnarray} \label{eq:real_up}
A^{(3,1)}_{[31]}(g_7^+) &=&
 g_{ew} \sqrt{\frac{\pi}{M_W \Gamma_W}} \,
\frac{\sqrt{2} \ \langle 46\rangle \, 
\langle 3|4+6|5]\, \langle 3|7-1|2 ]}
 {(s_{137} + M_W^2)\Delta_t \ \langle 17 \rangle \langle 37 \rangle} \, ,
\nonumber \\
A^{(3,1)}_{[31]}(g_7^-) &=&  
-g_{ew} \sqrt{\frac{\pi}{M_W \Gamma_W}} \,
\frac{\sqrt{2} \ \langle 46\rangle [12] \, [1|3+7|4+6|5] }
{(s_{137} + M_W^2)\Delta_t \ [17] [37]} \, ,
\end{eqnarray}
where $s_{137} = s_{13} + s_{17} - s_{37}$. For the lower-line
emission we obtain 
\begin{eqnarray} \label{eq:real_low}
A^{(3,1)}_{[42]}(g_7^+) &=&
-g_{ew} \sqrt{\frac{\pi}{M_W \Gamma_W}} \, \frac{\sqrt{2}  \langle 46 \rangle}
   {(s_{13} +M_W^2)\Delta_t}
\times
\nonumber \\
&& \ 
\bigg( \frac{\langle 3|4+6|5] \langle 4|7-2|1]}
           {\langle 27 \rangle \langle 47 \rangle} +
   \frac{[12]}{\langle 47 \rangle} 
   \frac{\langle 3|1+2|7] \langle 4|6|5] - 
               \mu_t^2 \langle 34 \rangle [57]}{\Delta_{t7}} 
\bigg) \,  ,
\nonumber
\\
A^{(3,1)}_{[42]}(g_7^-) &=& 
-g_{ew} \sqrt{\frac{\pi}{M_W \Gamma_W}} \,  \frac{ \sqrt{2} \ [12]}
   {(s_{13} + M_W^2) \Delta_{t7}}
\times
\nonumber \\
&& \ 
\bigg( \frac{\langle 3|1+2|5] \langle 6|4+7|2]}{[27][47]} -
       \frac{\langle 4 6 \rangle}{[27]}
       \frac{\langle 3|1|2] \langle 7|4+6|5] + \mu_t^2 \langle 37
         \rangle [25]} {\Delta_t}
\bigg) \, ,
\end{eqnarray}
with $\Delta_{t7}=(p_t+p_7)^2-\mu_t^2$. The complex mass, $\mu_t$, in
the numerator of \Eqn{eq:real_low}, which follows from the on-shell
matching condition, $p_t^2 = \mu_t^2$, guarantees that QCD Ward
identities are satisfied exactly. As already pointed out for the hard
virtual corrections computed in Section~\ref{sec:virtual}, the
matching condition needs to be satisfied only order-by-order in
$\delta$, and one could equally well replace $\mu_t$ with $m_t$ in the
numerator of \Eqn{eq:real_low}.  This would lead to a gauge-invariance
violation proportional to $\Gamma_t/m_t \sim \delta$, which is a
higher-order effect in our counting scheme.

Besides the process $u\,b \rightarrow d\,b\,l^+\,\nu_l\, g$, at order
$\delta^{1/2} \alpha_s$ the gluon-initiated processes $u\,g
\rightarrow d\, b\, l^+\,\nu_l\, \bar{b}$ and $g\, b \rightarrow d\,
b\, l^+\,\nu_l\, \bar{u}$ also contribute to the amplitude for
single-top production. The corresponding helicity amplitudes can
easily be obtained from the crossing of
\Eqns{eq:real_up}{eq:real_low}.

\subsection{Comparison of various approximations}

The results presented in this section have been obtained by making
kinematic approximations to the virtual and real corrections. Because
real and virtual corrections are infrared divergent, we have to make
sure that the approximations are consistent in that the real and
virtual soft and collinear singularities still cancel. Since, in
Section~\ref{sec:results}, we will be comparing our results to
previous calculations, it is instructive to discuss the cancellation of
infrared singularities and the relation between successive
approximations to $t$-channel single-top production.

The first approximation we consider is to treat the top as a stable
particle~\cite{Bordes:1994ki, Stelzer:1997ns, Harris:2002md,
  Sullivan:2004ie, Campbell:2009ss}. We will call this the stable-top
calculation and denote the process by $u\, b \to d\, t$. Within this
approximation, the renormalized virtual
corrections read
\begin{equation}
A^{{\rm virt}}_{u\, b \to d\, t} = A^{{\rm tree}}_{u\, b \to d\, t}
\left( \delta V_{13}^{H} + \delta V_{2t}^{H} 
+\frac{\delta m_t}{2 m_t}\right) + {\rm finite} \ ,
\label{Astable}
\end{equation}
where $A^{{\rm tree}}_{u\, b \to d\, t}$ is the corresponding tree-level
amplitude; $\delta V^H_{13}$, $\delta V^H_{2t}$ and $\delta m_t/m_t$
are given in \Eqns{eq:hard}{deltam}; and ``finite'' represents
non-singular terms that cannot be factorized in terms of the
leading-order amplitude. Thus, the virtual corrections in
\Eqn{Astable} are precisely the leading hard corrections of
diagrams~(c) and (d) in Figure~\ref{fig:1loop} without the top decay,
plus the renormalization of the external top-quark line, $\delta
Z_t/2=\delta m_t/(2 m_t)$. The corresponding infrared singularities are
cancelled, in the usual way, by the real corrections due to the process
$u\, b \to d\, t\, g$.

The next approximation we consider is the calculation presented in
Ref.~\cite{Campbell:2004ch}, which we call the on-shell calculation
and denote by $u\, b \stackrel{t}{\to} d\,b\, W$.  In this
computation, the production of an on-shell top quark is combined with
the decay of an on-shell top. This leads to additional singularities
in the virtual matrix element, which is given by
\begin{equation}
A^{{\rm virt}}_{u\, b \stackrel{t}{\to} d\,b\, W} = 
A^{{\rm tree}}_{u\, b \stackrel{t}{\to}  d\,b\, W}
\left( \delta V_{13}^{H} + \delta V_{2t}^{H}
+ \delta V_{4t}^{H} +\frac{\delta m_t}{m_t} \right) + {\rm finite} \, .
\label{Aonshell}
\end{equation}
The additional contribution, $\delta V^H_{4t}$ (given in \Eqn{eq:hard}),
corresponds to the one-loop correction to the top decay, while $\delta
m_t/m_t = 2 \times \delta m_t/(2 m_t)$ accounts for the
renormalization of the top-quark legs in the production \emph{and}
decay vertices. In order to match this with the real corrections, the
latter have to be treated with some care. In particular, the
interference between real gluon radiation from the production part of
the process with the decay part has to be neglected. However, real
corrections restricted to either the production or decay part are
taken into account and the corresponding singularities cancel those of
the virtual corrections, \Eqn{Aonshell}.

In the calculation presented in this paper we move from on-shell top
production to resonant top production, indicated by a star, $u\, b
\stackrel{t^*}{\to} d\,b\, W$.  As discussed in the previous section,
this involves a whole tower of additional contributions. However, to
the accuracy we are aiming for, the only additional contribution is due
to the (leading) soft part of the diagrams in Figure~\ref{fig:1loop},
given in \Eqn{eq:soft}. Thus, we have additional virtual
singularities, given in \Eqn{softall}, and we obtain
\begin{equation}
A^{{\rm virt}}_{u\, b \stackrel{t^*}{\to} d\,b\, W} = 
A^{{\rm tree}}_{u\, b \stackrel{t^*}{\to} d\,b\, W}
\left( \delta V_{13}^{H} + \delta V_{2t}^{H}
+ \delta V_{4t}^H+ \frac{\delta m_t}{m_t}
+ \delta V^S \right) + {\rm finite} \, ,
\label{Aresonant}
\end{equation}
with $A^{{\rm tree}} = A^{(3,0)}_{(-1)}$.  The new contribution,
$\delta V^S$, exactly cancels the single poles in \Eqn{Aonshell} due
to soft emission off a massive top-quark leg.  Thus, \Eqn{Aresonant}
contains only singularities corresponding to collinear or soft
emission off massless legs, as it should for a process with only
massless external states.  The additional poles in $\delta V^S$ are to
be cancelled by real singularities which essentially correspond to
those neglected in the process $u\, b \stackrel{t}{\to} d\,b\, W$,
i.e. the interference between real radiation from the production and
decay of the top. From a more formal point of view, one has to compute
the real matrix element squared of the process $u\, b \to d\,b\, W\,
g$ and expand it in $\delta$. Because we only need the leading term in
$\delta$, it is sufficient to consider the diagrams of
Figure~\ref{fig:real}. Taking the amplitude squared corresponding to
these diagrams, integrating over the phase space and expanding the
result in $\delta$ leads to real singularities that precisely cancel
those of the virtual corrections to the process $u\, b
\stackrel{t^*}{\to} d\,b\, W$.

We stress once more that the calculation presented here is only
meaningful if the top is nearly on-shell. If we are interested in the
process $u\, b \to d\,b\, W$ with no constraint on the invariant mass
of the final state particles, we have to compute the full virtual and
real corrections, i.e. also take into account the diagrams that have
been omitted in Figures~\ref{fig:1loop} and \ref{fig:real}. Needless
to say, such a computation is considerably more involved, but the
cancellation of real and virtual singularities works in a
straightforward way.

\section{Results}
\label{sec:results}

The results for the three approximations outlined in the previous
section, $u\, b \to d\, t$, $u\, b \stackrel{t}{\to} d\,b\, W$ and
$u\, b \stackrel{t^*}{\to} d\,b\, W$ have been implemented using Monte
Carlo integration, allowing us to calculate numerical values for both
cross sections and kinematical distributions.  In this section we
present a selection of these results and include comparisons to
existing results where available. For simplicity, throughout this 
section the cross sections calculated using the three approximations 
will be referred to as $\sigma^{{\rm prod}}$, $\sigma^{t}$ and $\sigma^{t^*}$ 
for the stable-top production, on-shell production followed by decay and 
resonant-top calculations respectively.

\subsection{Total cross sections}
\label{sec:xsections}

We begin by presenting a comparison of our results, obtained using the
three approximations, to those for the production of a single,
stable top quark as presented in Ref.~\cite{Campbell:2009ss}. We
compare the results for an LHC run at a centre of mass energy
$\sqrt{s} = 10$~TeV, and use MSTW2008 PDFs~\cite{Martin:2009iq} and
the corresponding strong coupling.  Renormalization and factorization
scales are set to be equal to a value of $m_t/2$.  The other input
parameters used for this calculation are shown in
Table~\ref{table:CFMTinputs}. In the top-quark propagator, we use the
tree-level decay width for the LO cross section and the
$\alpha_s$-corrected width for the NLO calculation.  This ensures
that, at leading order, we obtain agreement between the stable-top
production cross section and that of on-shell production plus decay,
after integration over the fully inclusive decay of the top quark and
$W$-boson, (i.e. $\sigma^{{\rm decay}}_0=\Gamma_t^{{\rm LO}}$).  The
results of our calculations using these parameters are shown in
Table~\ref{table:CFMTresults}.  We include the results for on-shell
production taken from Table~5 of Ref.~\cite{Campbell:2009ss} for ease
of comparison.
 
The agreement with the existing results at both leading and
next-to-leading order is very good when we perform the calculation for
the production of a stable top. When we also include the subsequent
decay of the on-shell top we still have good agreement at LO, but we
no longer agree at NLO.  This discrepancy is due to the use of the
improved narrow width approximation for the on-shell top.  We can see
from
\begin{equation}
\sigma^{t}_0 = 
\frac{\sigma^{{\rm prod}}_0 \sigma^{{\rm decay}}_0}
{\Gamma_t^{{\rm LO}}}= 
\frac{\sigma^{{\rm prod}}_0 \Gamma_t^{{\rm LO}}}
{\Gamma_t^{{\rm LO}}} = \sigma^{{\rm prod}}_0
\label{eq:LO} 
\end{equation}
that at LO the dependence on the decay width of the top quark
cancels. However, at NLO there is no full cancellation and we are left
with a residual dependence on the width,
\begin{equation}
\sigma^{t} = 
\frac{\sigma^{{\rm prod}}_0 \Gamma_t^{{\rm NLO}}+\sigma^{{\rm prod}}_1 
\Gamma_t^{{\rm LO}}}{\Gamma_t^{{\rm NLO}}} = \sigma^{{\rm prod}} + 
\sigma^{{\rm prod}}_1\frac{\Gamma_t^{{\rm LO}}-\Gamma_t^{{\rm NLO}}}
{\Gamma_t^{{\rm NLO}}}
\label{eq:NLO} \, .
\end{equation}
The difference between $\sigma^t$ and $\sigma^{{\rm prod}}$ is
formally a higher-order correction, since $\sigma_1^{{\rm prod}}
(\Gamma_t^{{\rm LO}}- \Gamma_t^{{\rm NLO}}) \sim \alpha_s^2$, but
leads to a visible numerical effect.  From the resonant top
calculation (last column in Table~\ref{table:CFMTresults}), it is
clear that taking into account the non-factorizable corrections causes
a small, but noticeable increase of the cross section at both LO and
NLO.

\begin{table}
  \begin{center}
  \begin{tabular}{lr}

  \hline
  \hline
  $m_t=172$~GeV & $\Gamma_W=2.05141$~GeV \\[2pt]
  $M_W=80.4$~GeV & $\Gamma_t^{{\rm LO}}=1.46893$~GeV \\[2pt]
  $\alpha_{ew}=0.03402$ & $\Gamma_t^{{\rm NLO}}=1.32464$~GeV \\
  \hline
  \hline

  \end{tabular}
  \end{center}
  \ccaption{}{Input parameters used for calculating the cross sections shown in 
    Table~\ref{table:CFMTresults}. \label{table:CFMTinputs}} 
 \end{table} 

\begin{table}
  \begin{center}
  \begin{tabular}{l|c|ccc}

  \hline
  \hline
  &  Ref.~\cite{Campbell:2009ss} & $\sigma^{{\rm prod}}$ & $\sigma^{t}$ 
& $\sigma^{t^*}$  \\
  \hline
  LO (pb) & 76.6 & 76.62(1) & 76.62(1) & 77.36(5) \\[2pt]
  NLO (pb) & 84.4 & 84.41(1) & 84.91(2) & 86.3(3) \\
  \hline
  \hline

  \end{tabular}
  \end{center}
  \ccaption{}{Comparison of total cross sections, calculated using our
    three methods, to those of  Campbell et al. (2009)
    \cite{Campbell:2009ss} at leading order (LO) and next-to-leading  
    order (NLO).\label{table:CFMTresults}}
\end{table} 

We now move on to discuss the comparison of our results to those in
Ref.~\cite{Campbell:2004ch} for the on-shell production of a single
top quark followed by its decay. We compare the results for an LHC run
with centre of mass energy $\sqrt{s}=14$~TeV. The input parameters can
be found in Table~IV of Ref.~\cite{Campbell:2004ch}. For this
comparison we set the renormalization and factorization scales equal
to $m_t$ and use the MRST2002 NLO PDFs~\cite{Martin:2002aw} and the
corresponding $\alpha_s$ value. The results are shown in
Table~\ref{table:CETresults}. The total cross section in the
stable-top production case is obtained by multiplying the production
cross section, $\sigma^{{\rm prod}}$, by the leading-order branching
ratio for the top-quark decay, $B_{t \rightarrow be\nu} = 0.1104$.

\begin{table}
  \begin{center}
  \begin{tabular}{l|cc|cc|cc}

  \hline
  \hline
  &  $\sigma^{{\rm prod}}_0 B_{t \rightarrow be\nu}$ & $\sigma^{{\rm prod}} B_{t \rightarrow be\nu}$ 
  & $\sigma^{t}_0$ & $\sigma^{t}$ & $\sigma^{t^*}_0$ & $\sigma^{t^*}$ \\
  \hline
  Ref.~\cite{Campbell:2004ch} (pb) & 17.69(1) & 17.05(2) & 17.69(1) & 16.98(2) & N/A & N/A \\[2pt]
  Our results (pb) & 17.71(1) & 17.04(1) & 17.71(1) & 16.98(1) & 17.94(1) & 17.33(8) \\
  \hline
  \hline

  \end{tabular}
  \end{center}
  \ccaption{}{Comparison of total cross sections, calculated using our
    three methods, to those of  Campbell et al. (2004)
    \cite{Campbell:2004ch} at leading order ($\sigma_0$) and
    next-to-leading  order ($\sigma$).  \label{table:CETresults}}
\end{table}

\begin{table}
  \begin{center}
  \begin{tabular}{lr}

  \hline
  \hline
  $m_t=172$~GeV & $M_Z=91.2$~GeV \\[2pt]
  $M_W=80.4$~GeV & $p_T(J_b) > 20$~GeV \\[2pt]
  $\alpha_{ew}=0.03394$ & $p_T(e) > 25$~GeV \\[2pt]
  $\Gamma_W=2.14$~GeV & $\dirac{E_T} > 25$~GeV \\[2pt]
  $\Gamma_t^{{\rm NLO}}=1.32813$~GeV & $120 < m_{{\rm inv}} < 200$~GeV \\
  \hline
  \hline

  \end{tabular}
  \end{center}
  \ccaption{}{Input parameters used for calculating the cross sections shown in 
Table~\ref{table:xsec}.\label{table:parameters}}
\end{table} 
\begin{table}
  \begin{center}
  \begin{tabular}{l|cc}

  \hline
  \hline
  & $\sigma^{t}$ & $\sigma^{t^*}$  \\
  \hline
  LO (pb) & 2.6786(1) & 2.519(1) \\[2pt]
  NLO (pb) & 2.3079(1) & 2.227(4)  \\
  \hline
  \hline

  \end{tabular}
  \end{center}
  \ccaption{}{Comparison of leading order (LO) and next-to-leading order
    (NLO) total cross sections  for our minimal realistic
    setup. \label{table:xsec}} 
\end{table} 

As in the first comparison, we obtain good agreement between our results 
and the existing results for the stable-top production and on-shell production 
plus decay calculations. The inclusion of the non-factorizable
corrections again has the effect of increasing the cross
sections by a modest amount at both LO and NLO.

Finally, we look at the total cross section for our minimal realistic
setup, described in Section~\ref{sec:setup}. For illustration, jets
are constructed using a standard $k_\perp$ cluster algorithm with the
resolution parameter set to $D_{\rm res} = 0.7$, but any other jet
definition could also be used. We assume that we can always identify
$J_b$, the jet containing the $b$ quark.  We apply cuts on $p_T(J_b)$
and $p_T(e)$, the transverse momenta of $J_b$ and the positron
respectively, and on the transverse missing energy, $\dirac{E_T}$. In
addition, we require that the invariant top mass, which is defined as
\begin{equation}
 m_{{\rm inv}}=\sqrt{\left( p(J_b)+p(e)+p(\nu) \right)^2},
\label{eq:minv}
\end{equation}
is close to the top-quark resonance. We perform the calculation for an
LHC run with $\sqrt{s}=7$~TeV and use the MSTW2008 NLO PDFs. The
renormalization and factorization scales are set to $m_t/2$. A full
list of the input parameters and cuts can be found in
Table~\ref{table:parameters}. We use NLO PDFs and the NLO top decay
width at all times to accentuate the off-shell effects by ensuring
that the corrections are not coming from the change of PDF or
width. As we now apply cuts to the decay products of the top quark, it
is no longer possible to include the stable-top production
calculation, as this requires fully inclusive decays. 

The results for the total cross section are shown in
Table~\ref{table:xsec}.  With the introduction of the cuts, the
effects of the non-factorizable corrections become more
pronounced. Instead of increasing the cross sections, as was the case
in the earlier comparisons (Tables~\ref{table:CFMTresults} and
\ref{table:CETresults}), the cross sections are now decreased by the
inclusion of these corrections. The relative size of the
off-shell contributions also increases from $\sim 1\%$ to a few
percent.

\subsection{Distributions}
\label{sec:distributions}

Partial results for kinematical distributions computed with the method
presented here have been shown in Ref.~\cite{Falgari:2010nt}. In this
section we will consider the minimal realistic setup described in the
previous subsection, and use the parameters and cuts given in
Table~\ref{table:parameters}.

\begin{figure}[t]
  \epsfxsize=12cm \centerline{\epsffile{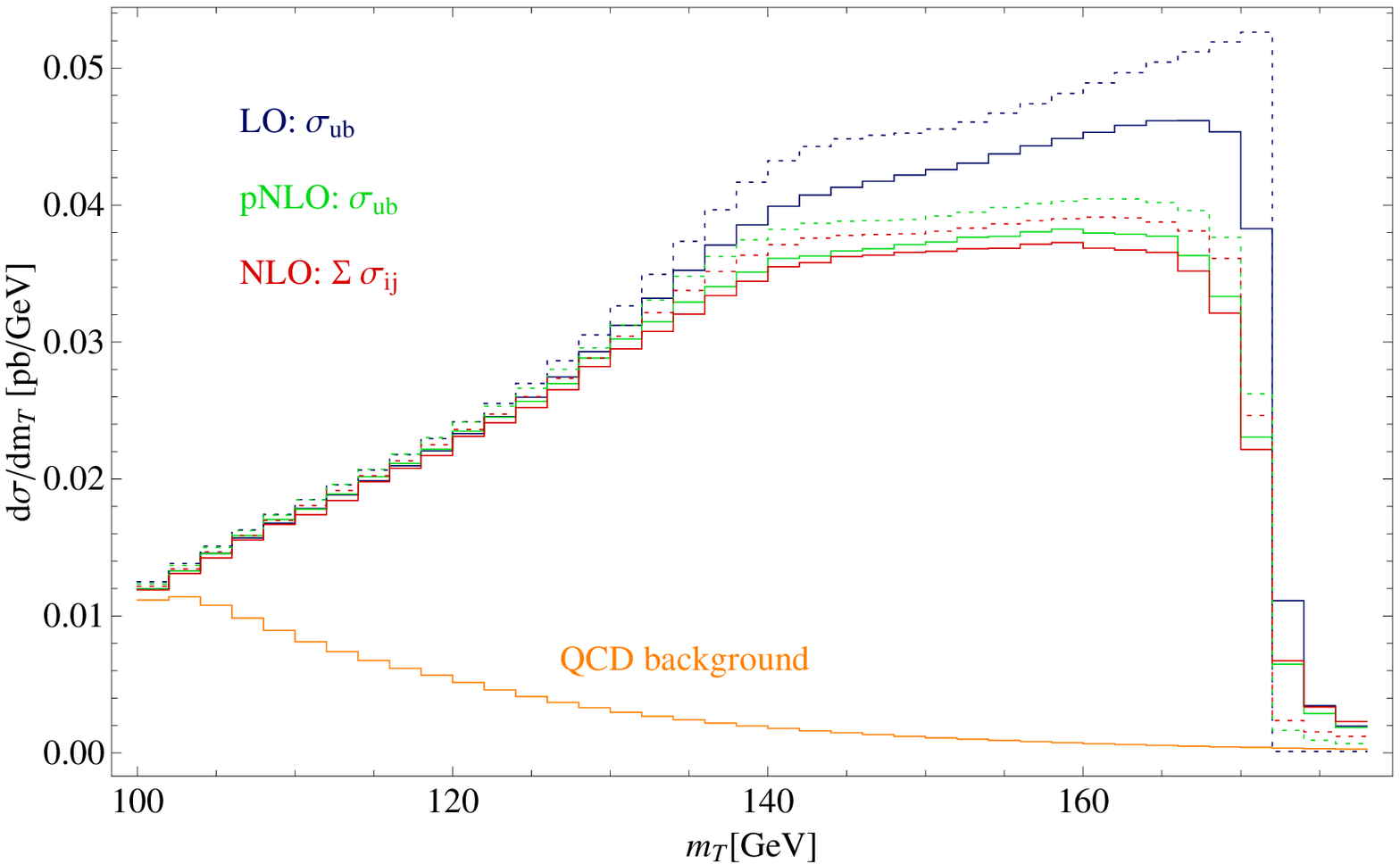}}
  \ccaption{}{Transverse top mass distribution for on-shell (dashed
    lines) and resonant (solid lines) top-quark production. LO results
    are shown in red, pNLO results (see text for explanation) in green
    and full NLO results in red. The orange line shows the subleading
    QCD contribution.\label{fig:Mtr}}
\end{figure}

The first distribution we present is that of the transverse mass of
the top, defined as
\begin{equation}
\label{mTdef}
m^2_{\rm T} = |p_T(J_b)|^2+ |p_T(e)|^2+ |p_T(\nu)|^2  - 
\left( \vec{p}_T(J_b) + \vec{p}_T(e) + \vec{p}_T(\nu)\right)^2 .
\end{equation}
We will compare on-shell ($u\, b \stackrel{t}{\to} d\,b\, W$) and
resonant ($u\, b \stackrel{t^*}{\to} d\,b\, W$) top-quark production,
shown as dashed and solid lines respectively in
Figure~\ref{fig:Mtr}. To assess the importance of higher-order
corrections, we will consider LO results (blue); full NLO results
(red), including also partonic processes with gluons in the initial
state; and partial NLO results (pNLO, green), including only one-loop
corrections to partonic processes that are present at LO.  Overall,
the differences between the on-shell and resonant results are
small. However, the off-shell effects do significantly change the
shape of the distribution near the boundary $m_{\rm T} = m_t$. As
expected, the LO on-shell distribution shows a sharp edge at this
boundary. The on-shell pNLO and NLO results have a contribution for
$m_{\rm T} > m_t$, because the $b$ jet, $J_b$, can contain gluon
radiation, however, this contribution is very small. In the resonant
calculation, we get a contribution in the region $m_{\rm T} > m_t$ even
at LO. This contribution is reduced at NLO but is still significantly
larger than in the on-shell result.

It can be seen from \Eqn{treeM} that there are also terms of higher order in
$\delta$ in the squared tree-level matrix element. First, there are
the interference terms $g_{ew}^6\, 2\, {\rm Re} ( A^{(3,0)}_{(-1)}\,
[A^{(3,0)}_{(0)}]^*)$; second there are the terms proportional to
$g_{ew}^2\, g_s^4\, |A^{(1,2)}|^2$, commonly referred to as the QCD
background; and, finally, there are subleading electroweak corrections
$g_{ew}^6\, |A^{(3,0)}_{(0)}|^2$. The former two are ${\cal
  O}(\delta^2)$, whereas the latter is ${\cal O}(\delta^3)$.  Therefore,
parametrically, these terms are beyond our NLO approximation. However,
they are a subset of subleading corrections that are very easy to
compute and it is useful to compare their numerical importance with
respect to the terms of ${\cal O}(\delta^{3/2})$. Thus, the QCD
background is shown separately in Figure~\ref{fig:Mtr}. This
contribution is actually important in the region $m_{\rm T} \ll m_t$,
but is insignificant when $m_{\rm T} \sim m_t$. The subleading
electroweak terms of ${\cal O}(\delta^2)$ and ${\cal O}(\delta^3)$ are
very small, and on the scale of Figure~\ref{fig:Mtr} they would show
simply as a straight line along the bottom of the plot.

\begin{figure}[t]
  \epsfxsize=12cm \centerline{\epsffile{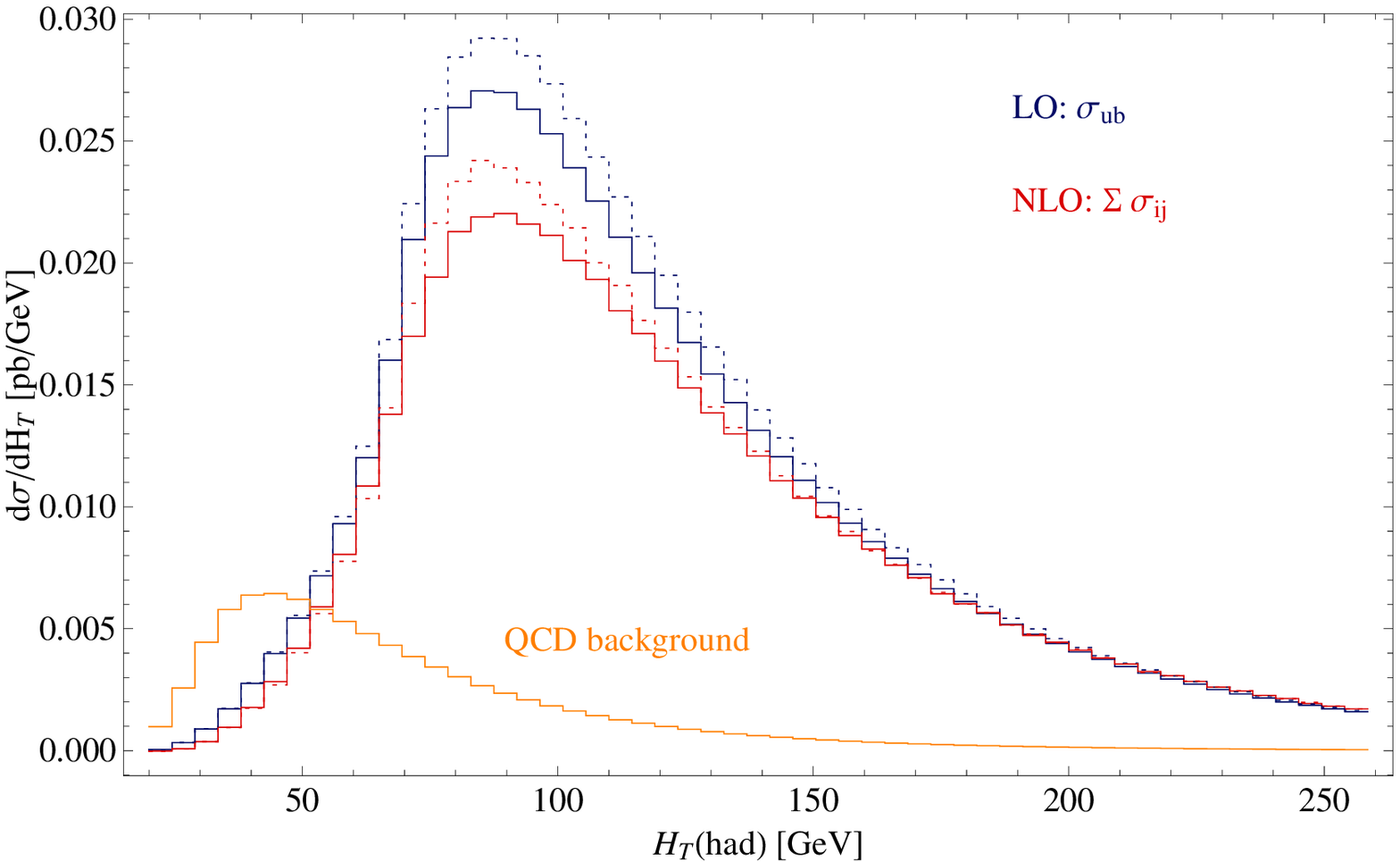}}
  \ccaption{}{$H_T({\rm had})$ distribution for on-shell (dashed) and
    resonant (solid) top-quark production at LO (blue) and NLO
    (red). The orange line shows the subleading QCD contribution.
   \label{fig:HT}}
\end{figure}

We now move on to look at our second distribution, the sum of the
hadronic transverse momenta, defined as
\begin{equation}
H_{\rm T}({\rm had}) = |p_T(J_b)| + |p_T(J_l)|\ ,
\label{eq:HT}
\end{equation}
where $J_l$ is the  (non $b$) jet with the largest transverse momentum.
The results are shown in Figure~\ref{fig:HT}. In this case, the
corrections due to the off-shellness of the top are relatively
large. The on-shell and resonant results differ by up to 10\% in some
bins. On average, the difference is approximately 3--4\%, in line with the
results presented in Table~\ref{table:xsec}. We have studied several
similar distributions and the effect of the off-shell corrections is
typically somewhat smaller than for $H_{\rm T}({\rm had})$. 
Again, the QCD background is separately shown in
Figure~\ref{fig:HT}. As for the transverse mass, it is
important only at the edge of the distribution, and the subleading
electroweak effects are again too small to be shown.

\begin{figure}[t]
   \epsfxsize=12cm
   \centerline{\epsffile{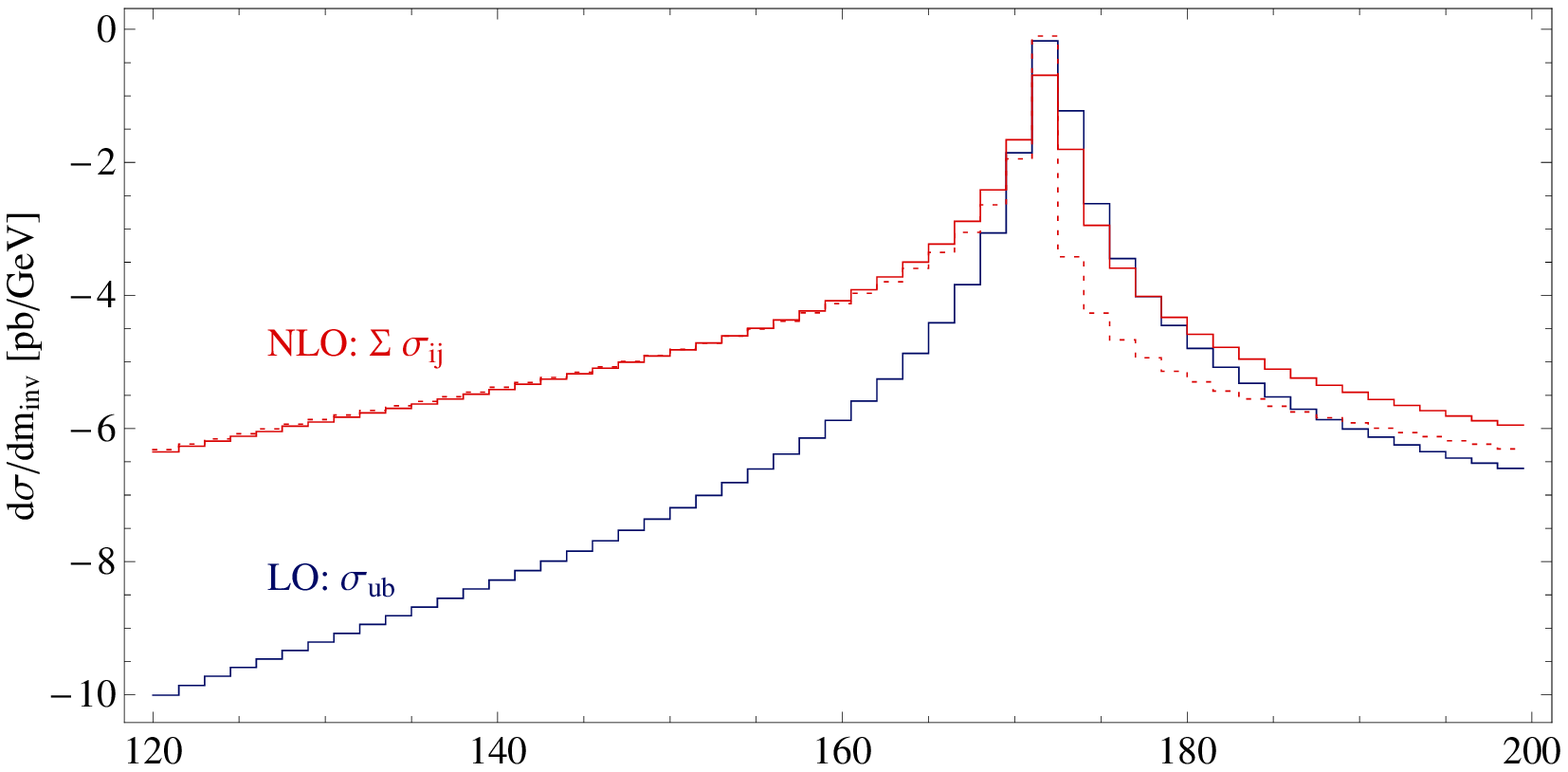}}
   \epsfxsize=12cm
   \centerline{\epsffile{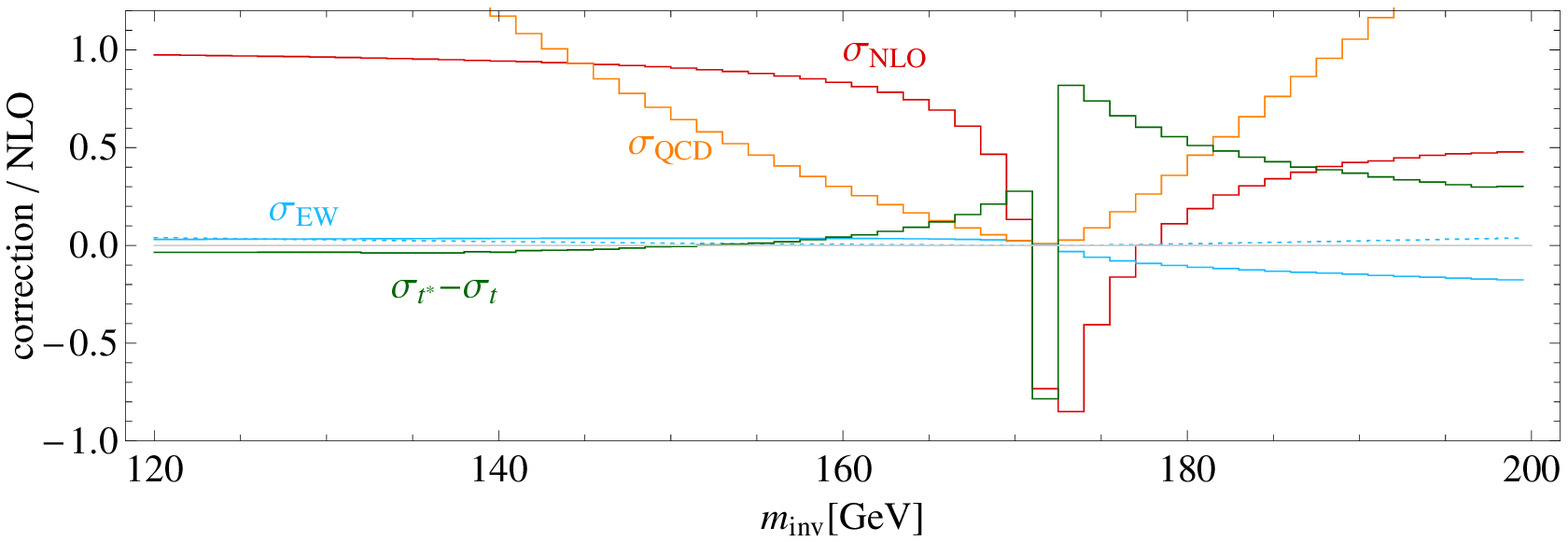}}
 \ccaption{}{Upper panel: Invariant mass distribution for LO (blue)
   and NLO (red) resonant top production, compared to NLO (red,
   dashed) on-shell top production. Lower panel: Ratios of various
   corrections (full NLO correction (red), QCD background (orange),
   off-shell corrections (green), subleading electroweak corrections
   (azure)) to the full NLO result.
   \label{fig:Minv}}
\end{figure}

Finally, we turn to the invariant mass of the top, as defined in
\Eqn{eq:minv}. The upper panel of Figure~\ref{fig:Minv} shows the
result at leading order for our resonant calculation, along with the
next-to-leading order results for both the on-shell and resonant
calculations. It should be noted that the leading order on-shell
production and decay distribution is a delta function centred at the
top mass and so would appear only as a single point at $m_t$ in the
figure. Therefore, it has been omitted.

We can see from the figure that the inclusion of NLO effects causes a
deviation from the Breit-Wigner shape of the distribution. We also
note that there is a visible difference between the NLO distributions
obtained via the two methods.  This difference is particularly
noticeable at invariant mass values greater than $m_t$.

Considering the shape of the invariant mass distribution also helps us
to understand why the inclusion of off-shell corrections increases the
total cross section when no cuts are applied but decreases it when we
apply some cuts. At LO, the invariant mass distribution has a
delta-spike shape in the on-shell case but a Breit-Wigner shape in the
resonant case. Taking into account NLO corrections, this picture is
modified somewhat, but comparing the on-shell and resonant
distributions, the former still has a more prominent peak at $m_{\rm
  inv} = m_t$, whereas the latter is larger in the region $m_{\rm inv}
> m_t$. Thus, there are two competing effects. If the cut on $m_{\rm
  inv}$ is mild enough, the increased contribution of the resonant
result for $m_{\rm inv} > m_t$ outweighs the larger peak of the
on-shell result, whereas for a tight cut on $m_{\rm inv}$, the
on-shell result is larger.

Note, however, that we must apply a cut to the invariant
mass, otherwise our assumption that the top quark is close to
resonance is no longer valid and the $\delta$-counting no longer
applies. As mentioned before, if we take our power counting at face
value we would expect our approximation to work in a window of $m_{\rm
  inv}-m_t\sim\Gamma_t$. However, the effective theory actually works
in a considerably larger window. To illustrate this, let us consider
the lower panel of Figure~\ref{fig:Minv}, where we compare the
relative importance of the various corrections to the full NLO
resonant distribution. The NLO corrections, $\sigma_{\rm NLO}$ (shown
in red), including all corrections of ${\cal O}(\delta^{3/2})$, are
dominant in the vicinity of $m_{\rm inv} \simeq m_t$, as
expected. They are larger than the ${\cal O}(\delta^2)$ QCD
corrections, $\sigma_{\rm QCD}$ (shown in orange), and the subleading
electroweak corrections, $\sigma_{\rm EW}$ (shown in azure). However,
away from resonance the formally subleading QCD corrections actually
become numerically more important than the formally leading
corrections. This is a clear indication that our power counting is no
longer applicable in this region and, therefore, the effective
theory breaks down. A similar point can be made in the case of the
subleading electroweak corrections, which have been divided into
${\cal O}(\delta^2)$ contributions (shown as solid azure line) and
${\cal O}(\delta^3)$ contributions (shown as dashed azure line).
These corrections are much smaller than our NLO corrections. In the
resonance region, the ${\cal O}(\delta^2)$ corrections are larger than
the ${\cal O}(\delta^3)$ corrections, but for $m_{\rm inv} \lesssim
130$~GeV this is no longer true. Once more, this indicates the
limitations of our power counting in this region. We should also
mention that for $m_{\rm inv} \lesssim 160$~GeV the NLO corrections 
are huge compared to the LO result and, therefore, our result is
not reliable. Finally, the off-shell effects, defined as the
difference between the full NLO results for the resonant and on-shell
calculations, are shown in green in the lower panel of
Figure~\ref{fig:Minv}. These effects are relevant near and above the
resonance region, but are very small below resonance.

We stress that the distributions presented here are only a sample of
the types of distribution that could be calculated. In principle, any
infrared safe quantity with arbitrary cuts on the final state
particles and jets could be easily computed.

\section{Conclusion and outlook}
\label{sec:conclusion}

In this work we have presented a method which allows the inclusion of
off-shell effects in resonant-particle production with a minimal
amount of computation. The method is based on a simultaneous expansion
of the cross section in the couplings and the small kinematic variable
$\Delta_t/m_t \simeq \Gamma_t/m_t$. It has been applied to
$t$-channel single-top production at the LHC. The calculation includes
the first non-trivial corrections to the narrow-width approximation,
corresponding to production-decay interference terms, and generalizes
earlier results of one-loop corrections to $t$-channel single-top
production.

Generally speaking, off-shell effects are small for inclusive
quantities. For the total cross section, for example, we find an
effect of the order of 1\%.  However, depending on the cuts applied,
off-shell effects can be sizeable. For most distributions we have
considered, the off-shell effects amount to a few percent of the LO
result, reaching up to 10\% in more extreme cases, such as the $H_{\rm
  T}({\rm had})$ distribution defined in \Eqn{eq:HT}. In particular,
they can significantly change the shape of distributions near
phase-space boundaries related to off-shell effects, such as the edge
for the transverse mass. This is, of course, not surprising since
sharp edges in distributions are usually related to having particles
on-shell. Thus, allowing the top quark to become slightly off-shell
can have a large impact in this region.

The calculation presented here can also be seen as a proof of the
workability of the effective-theory method. As we have shown in this
paper, the inclusion of the leading off-shell effects is relatively
straightforward, requiring in principle only the calculation of simple
soft corrections, since the hard part of loop integrals can be easily
related to results for on-shell production and decay of the massive
particle.  Furthermore, the computation of real corrections requires
only minor modifications to the standard subtraction procedure. Given
the generality of the effective-theory approach, the method can be
easily applied to other processes of phenomenological interest at the
LHC.  One such process is clearly represented by top-quark pair
production, which will be intensively exploited for measurements of
the top-quark properties, and whose study we reserve for future
publications.

\vspace*{0.5em}
\noindent
\subsection*{Acknowledgements}
We thank F.~Giannuzzi for comments on the manuscript.  The work of
P.F. is supported in part by the grant ``Premio Morelli-Rotary 2009''
of the Rotary Club Bergamo.  The work of P.M. is supported by an STFC
studentship.

\end{document}